\documentclass[journal]{IEEEtran}
\usepackage{amsmath,amsfonts}
\usepackage{algorithmic}
\usepackage{algorithm}
\usepackage{array}
\usepackage[caption=false,font=normalsize,labelfont=sf,textfont=sf]{subfig}
\usepackage{textcomp}
\usepackage{stfloats}
\usepackage{url}
\usepackage{verbatim}
\usepackage{graphicx}
\usepackage{cite}
\usepackage{balance}
\usepackage{wasysym}
\usepackage{siunitx}
\usepackage{color, xcolor}
\usepackage{tikz,xcolor}
\usepackage[hidelinks,bookmarks=false]{hyperref} 
\hypersetup{hidelinks,
	colorlinks=true,
	allcolors=black,
	pdfstartview=Fit,
	breaklinks=true}
	
\definecolor{lime}{HTML}{A6CE39}
\DeclareRobustCommand{\orcidicon}{
\begin{tikzpicture}
\draw[lime, fill=lime] (0,0)
circle[radius=0.16]
node[white]{{\fontfamily{qag}\selectfont \tiny \.{I}D}};
\end{tikzpicture}
\hspace{-2mm}
}
\foreach \x in {A, ..., Z}{%
\expandafter\xdef\csname orcid\x\endcsname{\noexpand\href{https://orcid.org/\csname orcidauthor\x\endcsname}{\noexpand\orcidicon}}
}
\graphicspath{{./figure/}}    

\begin{document}

\bstctlcite{IEEEexample:BSTcontrol}
\title{RHS-TRNG: A Resilient High-Speed True Random Number Generator Based on STT-MTJ Device}

\author{Siqing Fu$^1$\hspace{-1.5mm}\orcidA{}, Tiejun Li$^1$, Chunyuan Zhang\hspace{-1.5mm}\orcidD, Hanqing Li,\\ Sheng Ma\hspace{-1.5mm}\orcidE, Jianmin Zhang, Ruiyi Zhang and Lizhou Wu$^{\ast}$\hspace{-1.5mm}\orcidC

    \thanks{$^1$These authors contributed equally to this work.}
    \thanks{Manuscript received 12 February  2023; revised 17 May 2023 and 6 July 2023; accepted 19 July 2023. This work was supported by the State Key Laboratory of High Performance Computing Foundation (202201-13), Science and Technology on Parallel and Distributed Processing Laboratory Foundation (WDZC20205500115), National Key R\&D Project 2021YFB0300300, the NSFC 62172430, the NSF of Hunan Province 2021JJ10052, and the STIP of Hunan Province 2022RC3065.($^{\ast}$Corresponding author: Lizhou Wu.)}

    \thanks{Siqing Fu, Tiejun Li, Chunyuan Zhang, Hanqing Li, Sheng Ma, Jianmin Zhang, and Lizhou Wu are with the College of Computer Science and Technology, National University of Defense Technology, Changsha 410073, China (e-mail: \{fusiqingnudt, tjli, cyzhang, lihanqing23013, masheng, jmzhang, lizhou.wu\}@nudt.edu.cn).}%nudt

    \thanks{Ruiyi Zhang is with the Division of Science and Technology, Beijing Normal University - Hong Kong Baptist University United International College, Zhuhai 519087, China (e-mail: q030018105@mail.uic.edu.cn).}%uic
}

% <-this % stops a space

%\markboth{IEEE TRANSACTIONS ON VERY LARGE SCALE INTEGRATION (VLSI) SYSTEM,  Final Manuscript, July~2023.}%
%{Fu \MakeLowercase{\textit{et al.}}: RHS-TRNG: A Resilient High-Speed True Random Number Generator Based on Magnetic Tunnel Junction.}


\maketitle

\begin{abstract}
High-quality  random numbers are very critical to many fields such as cryptography, finance, and scientific  simulation, which calls for the design of reliable true random number generators (TRNGs). Limited by entropy source, throughput, reliability, and system integration, existing TRNG designs are difficult to be deployed in real computing systems to greatly accelerate target applications. This study proposes a TRNG circuit named RHS-TRNG based on spin-transfer torque magnetic tunnel junction (STT-MTJ).  RHS-TRNG generates resilient and  high-speed random bit sequences exploiting the stochastic switching characteristics of STT-MTJ. By circuit/system co-design, we integrate
RHS-TRNG  into a RISC-V processor as an acceleration component, which is driven by customized random number generation instructions. Our experimental results show that a single cell of RHS-TRNG has a random bit generation speed of up to 303 Mb/s, which is the highest among  existing MTJ-based TRNGs. Higher throughput can be achieved by exploiting cell-level parallelism. RHS-TRNG also shows strong resilience against PVT variations thanks to our designs using bidirectional switching currents and  dual generator units. In addition, our system evaluation results using gem5 simulator suggest that the system equipped with  RHS-TRNG  can achieve 3.4–12x higher performance in speeding up option pricing programs than software implementations of random number generation.

\end{abstract}


\vspace*{-0.8\baselineskip}  
\section{Introduction}

\IEEEPARstart{W}{ith} the proliferation of semiconductor products, random numbers play an increasingly vital role in many fields such as  cryptography, computational finance, scientific simulation, artificial intelligence(AI), and stochastic computing \cite{gonzalez2017encryption, abboud2020surprising, jia2019spinbis}.
Existing computing systems mainly rely on   pseudo-random number generators (PRNGs)  to generate random numbers. However, this software-based method compromises generation speed and quality because of its predictable and periodic characteristics; it may even open  doors to potential attacks that  compromise keys, intercept data, and ultimately hack devices and  communication channels. As an alternative, true random number generators (TRNGs)  are able to produce random number sequences that are truly uniformly distributed and unpredictable by sampling physically random processes \cite{Plumstead1982}. In addition, TRNGs are typically implemented in hardware, thus making them much faster than PRNGs in generating random bit stream. Therefore,  it is crucial to design a quality high-speed TRNG that can be integrated into computing systems to accelerate and secure critical applications.

Hardware-based TRNG designs rely on random physical phenomena as entropy sources, which are typically obtained from existing commodity circuits such as DRAM or from dedicated CMOS designs. In the former, random phenomena that can be utilized include DRAM retention failures\cite{hashemian2015robust,sutar2017d}, start-up value variations\cite{eckert2017drng}, or random reads caused by illegal DRAM commands\cite{pyo2009dram,kim2019d}. However, these entropy sources are mostly slow physical processes, making them difficult for TRNG designs to meet the throughput and latency requirements of some applications. The latter, such as differential ring oscillators\cite{kim2019,Choi2021} and metastability\cite{suresh2012,li2017}, rely on CMOS circuit thermal noise or oscillator jitter. These CMOS-based  TRNGs also have some drawbacks such as low generation speed\cite{liu2011,kim2019}. Additionally, the area and power consumption of CMOS-based TRNG designs are also unsatisfactory.

As semiconductor technologies evolve, some emerging post-CMOS devices with lower power consumption and smaller areas than transistors can provide new solutions for TRNG designs. For example, magnetic tunnel junction (MTJ) as a promising Spintronic device has been widely studied over the past decades, and its random switching behavior driven by spin-transfer torque (STT) provide opportunities for designing high-quality TRNGs. More specifically, the switching process of STT-MTJ between  its two magnetic states is intrinsically stochastic under the influence of thermal fluctuation, thus making it a perfect entropy source \cite{stohr2006magnetism}.  There exist some works on MTJ-based TRNG, but the performance of MTJ has not been fully exploited\cite{wang2016novel,amirany2020true,perach2019asynchronous,vatajelu2018high, qu2017true, qu2018variation,morsali2022process}. Examining the prior art reveals the following four drawbacks: 1) The random bit generation latency is high, reaching for example \SI{80}{\nano\second} in \cite{perach2019asynchronous};
2) lifetime is limited by the high reset current; 3) random number quality is susceptible to process/voltage/temperature (PVT) variations; 4) system integration and acceleration effects on applications are unclear.

In this paper, we design a high-speed STT-MTJ-based TRNG  which offers high resilience to PVT variations. It can be integrated into computing systems as an acceleration component with customized instructions to produce high-quality random bit stream for performance/security-critical applications.
In our RHS-TRNG design, the number of write operations in a random bit generation cycle is reduced from two to one by feedback control, which reduces the  generation latency. Eliminating the reset operation also greatly boosts the TRNG lifetime since the high reset current no longer flows through the underlying STT-MTJ device.
In addition, we leverage two mechanisms: 1)  bidirectional switching currents and 2) dual generator units to equip RHS-TRNG with resilience capability to PVT variations. Finally, we integrate RHS-TRNG into the instruction pipeline of RISC-V cores and design three custom instructions to drive the TRNG acceleration component. To evaluate the performance, power and area of RHS-TRNG, we have implemented its circuit and layout in the Cadence Virtuoso tool. Simulation results shows that the latency is \SI{3.3}{ns/bit}, power is  2.65-\SI{5.3}{pJ/bit}, and area is 14.5-\SI{24.29}{\micro\meter\squared/bit}. By integrating multiple STT-MTJ cells, the circuit can achieve better performance in power consumption and area per bit.
To evaluate the hardware acceleration effect on target applications, we have simulated our system design in gem5 and used Monte Carlo option pricing program as a test benchmark. The experimental results show that by integrating RHS-TRNG, we can obtain more than 3.4x speedup when running Monte Carlo option pricing program when compared to generating random numbers using software methods.

In summary, the main contributions of this paper are listed as follows: 
\begin{itemize}
\item We design a high-speed two-phase TRNG circuit based on STT-MTJ. With feedback control, we avoid the reset phase in each random bit generation cycle to achieve the generation speed  of \SI{303}{Mb/s} for a single TRNG cell.

\item  We enhance the resilience of RHS-TRNG to tolerate PVT variations by  utilizing bidirectional write current and dual-generator XOR design. 

\item We propose circuit-system co-design for RHS-TRNG by customizing a RISC-V processor and instructions. Gem5 simulation results show 3.4-12x performance acceleration for Monte Carlo option pricing applications in comparison  to software implementations. 

\end{itemize}

The rest of the paper is organized as follows: Section.~\ref{sec2} introduces the fundamentals of  TRNG, MTJ device and its switching behavior. Section.~\ref{sec3} elaborates the motivation of this paper. Section.~\ref{sec4} presents the circuit and system co-design of RHS-TRNG. In Section.~\ref{sec5}, we present the experimental setups and results at both circuit and system levels. Section.~\ref{sec6} compares this work with the prior art, while Section.~\ref{secdis} discusses some valuable future research topics. Finally, Section.~\ref{sec7} concludes the paper.
 
\vspace*{-0.8\baselineskip} 
\section{Background}
\label{sec2}

\vspace*{-0.2\baselineskip} 
\subsection{True Random Number Generator}

Random number generators are typically classified into two categories: pseudo-random number generators (PRNGs) and  true random number generators (TRNGs).
PRNGs generate random sequences by algorithms that transform an internal state and calculates an output value upon request. Once the initial seed is set,  the next state only depends on the previous state.
As a result, pseudo-random sequences can be predictable and controllable; they are only mathematically consistent with a random distribution\cite{Plumstead1982}. In contrast, TRNGs generate random numbers by sampling random  physical phenomena such as thermal noise, electromagnetic behavior, and quantum behavior. Since these entropy sources are intrinsically non-deterministic,  the state of each  cycle in a TRNG cannot be predicted  even if all states are known when it runs. Therefore, it can produce a truly random bit stream to applications which have stringent requirements on the quality of random numbers.

Conventional TRNGs are typically implemented by leveraging  key features of CMOS circuits, e.g.,  ring oscillation (RO) \cite{kim2019,Choi2021} and metastability \cite{suresh2012,li2017}. However, these CMOS-based TRNG designs do not provide adequate performance. For example, the TRNG in \cite{liu2011} using time-dependent dielectric breakdown provides only \SI{0.011}{Mb/s} random sequences, and the speed of the differential OR-based TRNG in \cite{kim2019} is \SI{8.28}{Mb/s}.
To design faster TRNG hardware, researchers are looking for new technologies. 
Among the emerging devices, STT-MTJ has a smaller area and lower power consumption than CMOS transistors\cite{chun2012scaling}. Moreover, its operating principle involves inherent physical random processes, making it a promising option for designing TRNGs\cite{wang2016novel,amirany2020true,perach2019asynchronous,vatajelu2018high, qu2017true, qu2018variation, morsali2022process}.

\vspace*{-0.2\baselineskip} 
\subsection{Magnetic Tunnel Junction}

\begin{figure}[b]
\centering
\includegraphics[width=8.8cm]{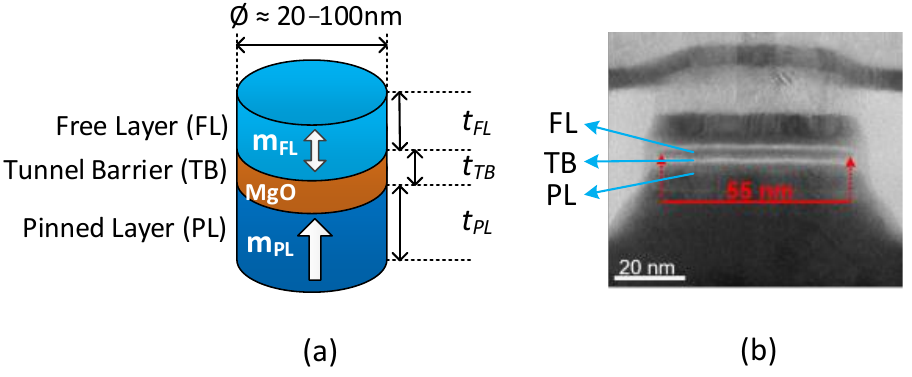}
\caption{Perpendicular MTJ device: (a) structure schematic and (b) cross-sectional TEM image.} 
\label{fig1}
\end{figure}

Magnetic tunnel junction (MTJ) is a widely used Spintronic device \cite{Gallagher,chen2012tunable,chen2012yoke}.
The core of an MTJ is a three-layer structure consisting of two ferromagnetic layers and one dielectric tunnel barrier (TB) layer sandwiched between them, as shown in Fig.~\ref{fig1}.a. The bottom ferromagnetic layer is called  pinned layer (PL), whose magnetization is fixed along the MTJ's easy axis \cite{Ikeda2007}. The top ferromagnetic layer is called free layer (FL); its magnetization is either in parallel (P) or anti-parallel (AP) to that of the PL \cite{yuasa2018materials}. Due to the tunneling magneto-resistance (TMR) effect \cite{mathon2001theory}, the AP and P magnetic states have different resistance values: the high-resistance value observed under the AP state ($R_{\mathrm{AP}}$) represents logic 1, while the low-resistance state under the P state ($R_{\mathrm{P}}$) represents logic 0. The difference between $R_{\mathrm{AP}}$ and $R_{\mathrm{P}}$ is commonly quantified using the TMR ratio, given by: $(R_{\mathrm{AP}}-R_{\mathrm{P}})/R_{\mathrm{P}}\times100\%$.

 The MTJ used in this paper is perpendicular MTJ composed of ferromagnetic layers with perpendicular magnetic anisotropy (PMA)\cite{park2015}. This type of MTJ has become the mainstream MTJ design in recent years, thanks to its small area and low power consumption. Fig.~\ref{fig1}.b shows a cross-sectional transmission electron microscopy (TEM) image of a $\diameter$ \SI{55}{\nano\meter} MTJ device fabricated at IMEC\cite{wu2019defect}. At present, Everspin Technologies has commercialized \SI{1}{\giga b} STT-MRAM chips with MTJs as persistent data-storing devices \cite{sun2021commercialization}. In addition, many foundries worldwide such as  TSMC, Samsung, and GlobalFoundries \cite{Chih,Lee2018,Naik2019} have claimed production service ready for cutting-edge  MTJ devices. 

\vspace*{-0.2\baselineskip} 
\subsection{STT Switching Stochasticity}
\label{Subsec:STTswitching_stochasticity}

\begin{figure}[t]
\centering
\includegraphics[width=7.3cm]{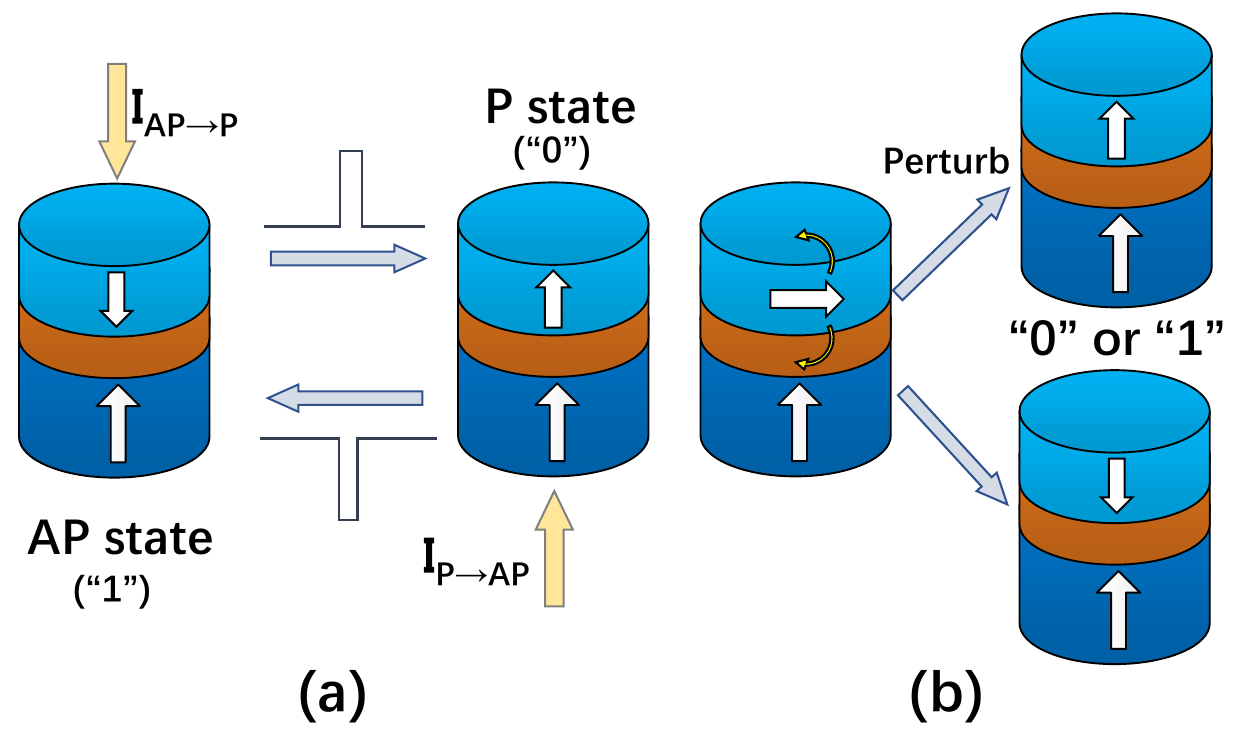}

\caption{(a) STT-MTJ's bipolar switching method between AP and P states. (b) Stochastic switching process to either ``0'' or ``1'' due to thermal perturbation.} 
\label{fig2}
\end{figure}

The MTJ state can be switched by several methods such as external magnetic field \cite{engel20054}, spin-transfer torque (STT) \cite{Gallagher2019IEDM}, and spin-orbit torque (SOT) \cite{Sato2020VLSI}.
An MTJ that uses STT effect to switch its binary state is called STT-MTJ. 
As shown in Fig.~\ref{fig2}.a, when a positive pulse is applied across the MTJ in AP state, it drives a current $I_{\mathrm{AP}\rightarrow\mathrm{P}}$ flowing perpendicularly from the FL to the PL. If the pulse's amplitude and width reach certain threshold, the FL's magnetization switches to the opposite direction, typically in 2-\SI{100}{\nano\second}. 
Similarly, a negative pulse can switch the MTJ from P to AP under the spin-polarized current $I_{\mathrm{P}\rightarrow\mathrm{AP}}$ when it exceeds the critical switching current. Note that STT-MTJ is a bipolar device, meaning that the polarity of the switching current determines   the  magnetization direction in the FL, which in turn determines the STT-MTJ's resistive state \cite{khvalkovskiy2013basic}.

Due to thermal fluctuation\cite{Wu2020survey}, the STT switching process of STT-MTJ is inherently stochastic for a given write pulse. The thermal fluctuation effect causes the magnetizations in the free layer (FL) and pinned layer (PL) to have a random, varying initial angle between them in each cycle, resulting in a random STT switching behavior, as shown in Fig.~\ref{fig2}.b. This produces a natural, cycle-to-cycle variance in the switching time. As a result, it is guaranteed that STT-MTJ, operating as an entropy source, can provide a random and independent bit sequence\cite{seki2011switching}.

\begin{figure}[t]
\centering
\includegraphics[width=8.8cm]{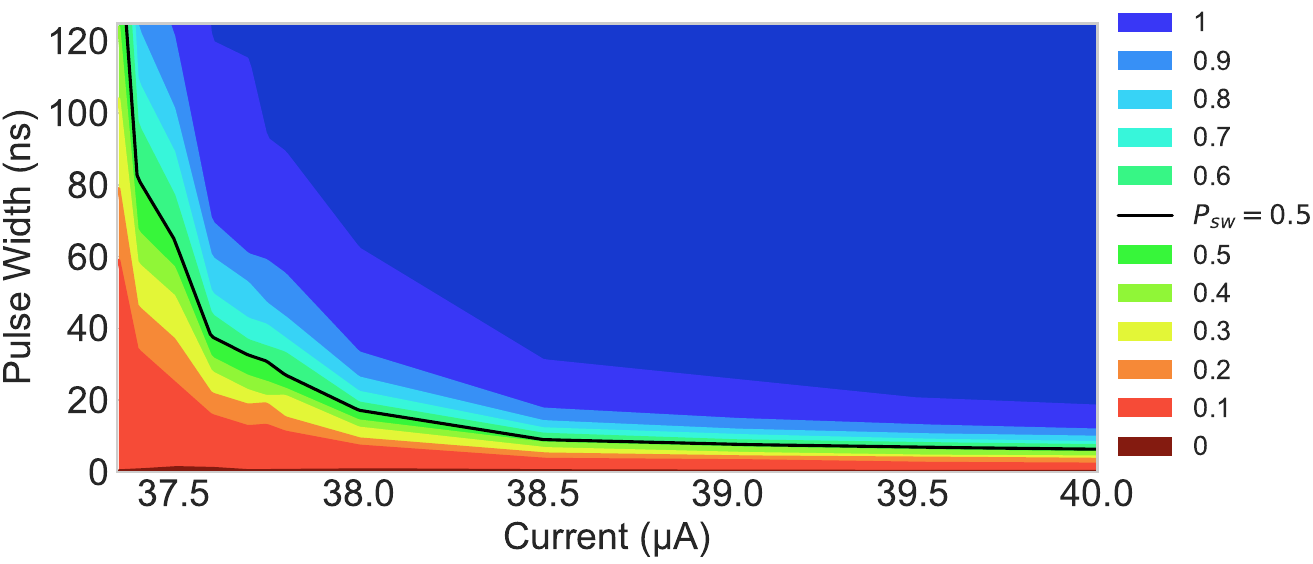}
\caption{The AP$\rightarrow$P switching probability of STT-MTJ  under different write current  and pulse width.} 
\label{fig3}
\end{figure}

As an example, Fig.~\ref{fig3} illustrates the simulated switching probability ($P_{\mathrm{sw}}$) from the AP to P state under various write pulse configurations. As shown in the figure, the switching probability increases with current and pulse width (from the lower-left to upper-right corner). The black line ($P_{\mathrm{sw}} = 0.5$) on the graph represents the operating points in which the switching probability is 50\%. This indicates that by tuning the write pulse to one of these operating points, the switching result  can be considered as a random variable $X$ which obeys the Bernoulli distribution: $X$$\sim$B($n$, 0.5).

In summary,  the switching stochasticity of STT-MTJ can serve  as a high-quality entropy source for TRNG design.

\vspace*{-0.8\baselineskip} 
\section{Motivation}
\label{sec3}

\vspace*{-0.2\baselineskip} 
\subsection{Application Demand for TRNGs}

In the era of information technology, the demand for high-quality random numbers is ubiquitous across various applications. Cryptography, for instance, requires good randomness in generating keys to ensure their security against attackers\cite{gonzalez2017encryption, qu2018variation}. This randomness is manifested in the form of long periodicity, non-linearity, unpredictability, among other factors. Similarly, in scientific simulations, the Monte Carlo method relies on high-quality random numbers to simulate random behavior in different processes\cite{dang2019, Ziegenhein2015}. Poor quality random bit streams can lead to degraded simulation confidence, and the generation rate of random numbers can also become a performance bottleneck in some Monte Carlo simulations. 
Apart from traditional scientific fields, emerging fields such as AI and stochastic computing also require high-quality random numbers. For instance, neural networks in AI applications are initialized with random numbers to break symmetry and enable faster convergence\cite{abboud2020surprising}.  Meanwhile, stochastic computing is an emerging computing paradigm that relies on random numbers for bit-wise operations\cite{jia2019spinbis}. Therefore, a high-quality TRNG is essential to ensure precise computation in these applications.

\vspace*{-0.2\baselineskip} 
\subsection{Limited System Integration with TRNGs}

Depending on application requirements and implementation technologies, TRNG hardware can be integrated into different parts of system architecture, from the pipeline, to cache, to memory, or simply as a peripheral connected to the system. However, the existing TRNG designs, regardless of their implementation technologies, rarely consider the effect of TRNG  integration on the target applications.

CMOS-based TRNG designs are typically evaluated using SPICE simulations \cite{kim2019,suresh2012}  or  FPGA prototypes \cite{Choi2021,li2017}. 
These works  evaluated TRNG hardware's energy consumption, randomness, and area, but they neglected system integration and its effects. For example, in \cite{Choi2021},  FPGAs have used to prototype cryptographic systems, but after  implementing TRNG on FPGA, they do not evaluate the effect of its integration into cryptographic systems.
MTJ-based TRNG designs are typically evaluated in circuit simulations using MTJ models\cite{wang2016novel,amirany2020true,perach2019asynchronous,vatajelu2018high, qu2017true, qu2018variation, morsali2022process}. These works have  fully analyzed the TRNG circuits composed of MTJ and CMOS, but rarely mentioned the higher level of the system. Even if the individual work evaluates system-level performance\cite{hou2022}, it does not consider the acceleration effect of TRNG on the benchmark. Another type of TRNG design is based on commercial DRAM\cite{hashemian2015robust,sutar2017d,eckert2017drng,olgun2021quac}.
Despite these works provide high feasibility of integrating into real systems, they require post processing and encroach memory bandwidth which is already a performance bottleneck in today's computing systems.

\begin{figure}[t]
	\centering
	\includegraphics[width=8cm]{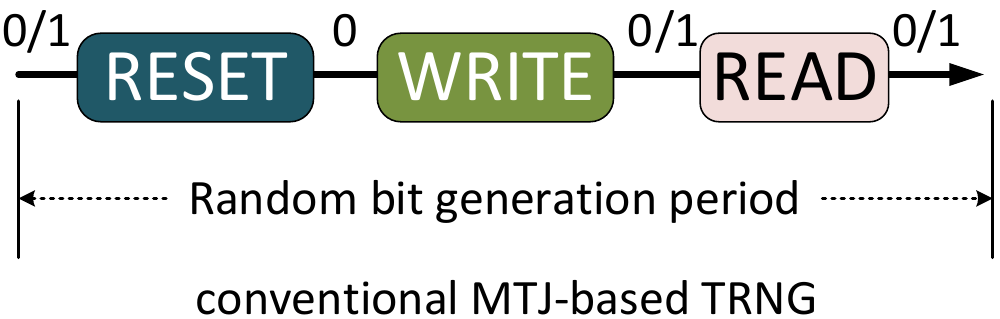}
	\caption{MTJ-based TRNG design with three phases: reset, write and read.} 
	\label{figmotiv}
\end{figure} 

\begin{figure}[t]
\centering
\includegraphics[width=8cm]{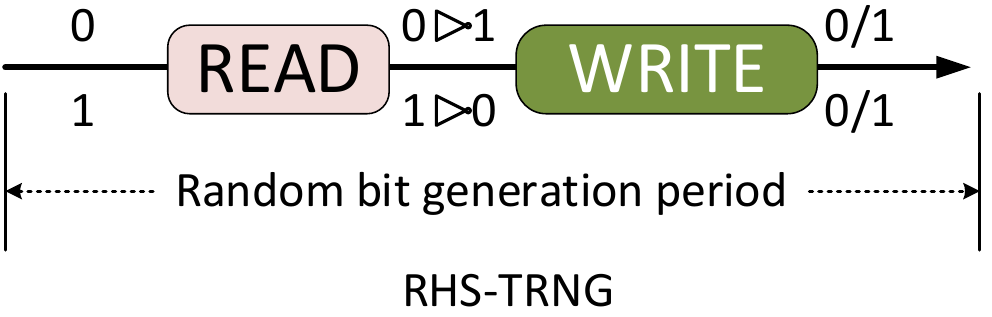}
\caption{ RHS-TRNG  with one cycle consisting of two phases: read and write.} 
\label{figtime_hr}
\end{figure}

\vspace*{-0.2\baselineskip} 
\subsection{Shortcomings of Prior MTJ-based TRNGs}
There exist several TRNG designs using STT-MTJ as an entropy source in the literature \cite{wang2016novel,amirany2020true,perach2019asynchronous,vatajelu2018high, qu2017true, qu2018variation, morsali2022process}. However, the performance of existing MTJ-based TRNG designs is significantly limited by their three-phase circuit design for generating a random bit in each cycle, as illustrated in Fig.~\ref{figmotiv}. The random bit generation process starts with the reset phase. This phase is designed to reset the MTJ to a fixed state (e.g., logic 0) from an initial state which can be either logic 0 or 1. This is achieved by applying  a large current going through the MTJ, causing it to switch to the P state with 100\% probability.  The second phase is random write. During this phase, the circuit applies a smaller write current to the MTJ in the opposite direction, which causes it to flip with a 50\% probability. As a result, the MTJ will be randomly set to logic 0 or 1 state. After the random write, the circuit goes through a read phase to read out the data  stored in the MTJ, i.e., the random bit output during this cycle.

Although the above traditional MTJ-based TRNG design is clear and feasible, it has the following three shortcomings. First, each cycle contains two  write operations: reset and random write. Given the fact that MTJ devices feature fast read and slow write, it limits the use of MTJ to design low-latency and high-throughput TRNGs. 
Second,  the reset phase requires a very large current flowing through the MTJ in order to guarantee 100\% switching probability (see Fig.~\ref{fig2}). This however may lead to  device breakdown, limiting the lifetime of the designed TRNG \cite{yoshida2009study}. 
Third, this simple TRNG  design is susceptible to process/voltage/temperature (PVT) variations. It has been demonstrated that process and voltage variations have a large impact on the MTJ's switching behavior \cite{Wu_TC2021}. In addition, an increase in the ambient temperature enables switching in both directions with reducing write voltage\cite{bi2012}.

While many traditional STT-MTJ TRNG designs require a reset phase, some recent advances in the field have led to  new possibilities. For instance, Choi et al.\cite{choi2014magnetic} proposed a TRNG that employs probability tracking with two pulse generators, eliminating the need for a reset phase and compensating for output probability fluctuations through counters and software-based real-time probability tracking. Similarly, Oosawa et al.\cite{oosawa2015design} introduced a digitally-controlled probability TRNG that does not rely on a reset phase and evaluated its ability to produce  random bit streams with stable probability distribution. However, these designs inevitably utilized post-processing circuits such as correction logics, D/A converters, and digital comparators to enhance the quality of randomness.

In summary, even though many MTJ-based TRNG designs are limited by their dependence on a reset phase, there have been  new opportunities due to recent breakthroughs. Our contributions to this advancement are threefold. First, we greatly improved the random number generation speed compared to the prior art. Second, in terms of addressing PVT variations, we avoided the use of post-processing circuits that would significantly increase power and area overheads. Third, we deployed our TRNG design in a RISC-V processor and corroborated its strength in accelerating specific applications.

\vspace*{-0.2\baselineskip} 
\subsection{Idea and Goal}
To ensure high-quality output bits and improve TRNG output rates to support applications that require acceleration, our ideas are: 1) the write overhead in each cycle of TRNG should be reduced as much as possible; 2) resilience to PVT variations should be obtained through dedicated circuit design to maintain the stability of the output probability of logic 0 and 1 in the generated bit streams; 3) TRNG hardware needs to be seamlessly integrated into existing computing systems as a random number acceleration component.
Motivated by these ideas, we aim to design a high-speed and resilient STT-MTJ-based TRNG and integrate it into existing computing platforms to greatly accelerate specific applications such as Monte Carlo simulations.

\begin{figure*}[t]
\centering
\includegraphics[width=14cm]{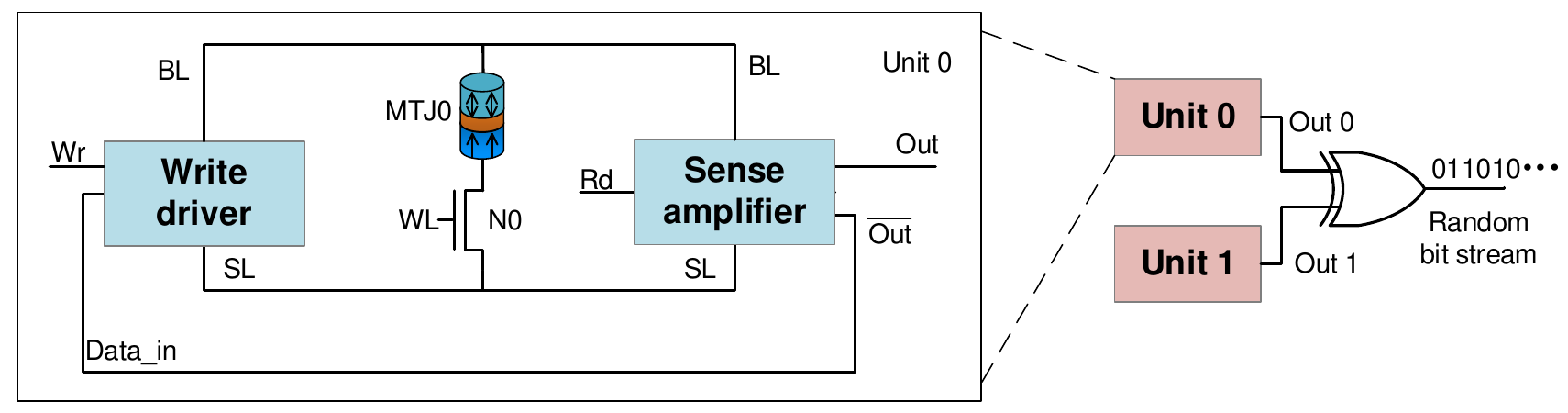}
\caption{Circuit design of RHS-TRNG.} 
\label{fig:overall_circuit_design}
\end{figure*}

\vspace*{-0.8\baselineskip} 
\section{Circuit-System Codesign}
\label{sec4}
In this section, we introduce the co-design of RHS-TRNG at  circuit and system levels. We first elaborate RHS-TRNG circuit design as well as its theoretical benefits. Thereafter, we detail the system and custom instruction design based on RISC-V ISA. 

\vspace*{-0.2\baselineskip} 
\subsection{RHS-TRNG Design}

\subsubsection{Design Philosophy}

Rather than relying on a single switching direction for generating random bit streams, we use current in both switching directions to control the random switch of the STT-MTJ. This enables us to eliminate the reset phase in each cycle of random bit generation, as depicted in Fig.~\ref{figtime_hr}.

At the beginning of each cycle, the MTJ's initial state (logic 0 or 1) is read out by a sense amplifier circuit, then the inverted value is fed to a write driver circuit. In the second phase, a random write operation is performed by the write driver; the write current direction depends on the data received from the previous read phase. This allows us to perform only one write operation in each cycle, which greatly reduces the time it takes to generate a random bit. Next, we will elaborate how to implement this two-phase TRNG design via VLSI circuits. 

\begin{figure}[t]
\centering
%\vspace{-20pt}
\includegraphics[width=8.8cm]{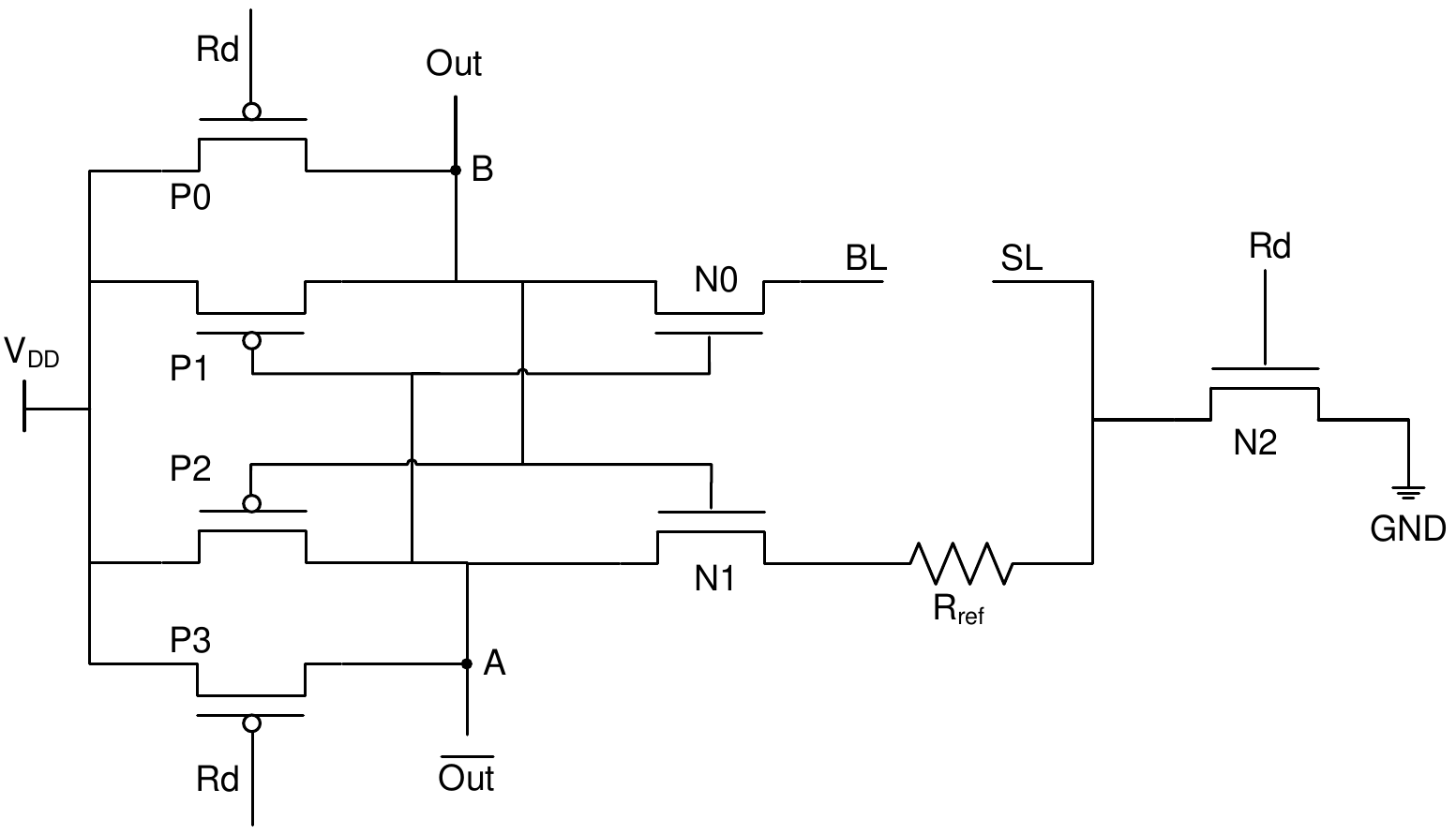}
\caption{Sense amplifier circuit design for RHS-TRNG.} 
\label{fig7}
\end{figure}

\subsubsection{Circuit Design}

The  circuit design of RHS-TRNG is shown in Fig.~\ref{fig:overall_circuit_design}. Each RHS-TRNG cell consists of two generator units and an XOR gate. 
The TRNG cell outputs a randon bitstream that is the XOR result of the two identical generator units. In each generator unit, an STT-MTJ device is connected in series with an NMOS transistor as a selector.  The on/off state of the NMOS is controlled by a word line (WL). The 1T-1MTJ structure is connected in parallel with a set of write driver and sense amplifier circuits  through a bit line (BL) and a source line (SL), which perform write and read operations on the STT-MTJ. When the ``$\mathrm{Rd}$'' signal is enabled, the sense amplifier can simultaneously read out the logic state (denoted as ``$\mathrm{Out}$'') of the MTJ and its inverted value ``$\overline{\mathrm{Out}}$''. The ``$\overline{\mathrm{Out}}$'' signal is then fed back to the write driver as an input ``$\mathrm{Data\_in}$''. When the ``$\mathrm{Wr}$'' signal is asserted, the write driver starts programming ``$\mathrm{Data\_in}$'' into the MTJ device by applying a pulse across the BL and SL. It generates a switching current whose direction is determined by the data to be programmed (i.e., ``$\mathrm{Data\_in}$'' ), as illustrated in Fig.~\ref{fig2}.

The sense amplifier circuit is shown in Fig.~\ref{fig7}. We use pre-charge sense amplifier design which compares the currents going through the MTJ cell under sensing and a fixed reference cell \cite{zhao2009high}. The entire read process is divided into the following three stages \cite{WuPhDThesis2021}. (1) \textbf{Pre-charge}. The ``Rd" signal is set to 0, which turns on PMOSs P0 and P3 and turns off NMOS N2. After the two nodes A and B are pre-charged to the same potential $V_{\mathrm{DD}}-V_{\mathrm{TH}}$, N0 and N1  are turned on, whereas P1 and P2 are turned off. (2) \textbf{Voltage development}.  The ``Rd" signal is set to 1, which controls P0 and P3 to be turned off and N2 to be turned on. The two nodes A and B  begin to discharge. The reference resistor is a fixed-resistance MTJ, so there is no area overhead of integrated resistors. Since the resistance value of the reference resistor is between $R_{\mathrm{P}}$ and $R_{\mathrm{AP}}$,  A and B will have different discharge rates. This results in a small voltage  quickly developed between A and B.
(3) \textbf{Voltage amplification}. Once the voltage reaches a threshold, it is quickly  amplified to the full swing by the cross-coupled inverters (P1, P2, N0, N1). Due to its fast read speed (hundreds of picoseconds)  and tiny sensing current, this sense amplifier circuit is ideal for designing high-speed and low-power TRNGs.

\begin{figure}[t]
\centering
\vspace{-20pt}
\includegraphics[width=8.8cm]{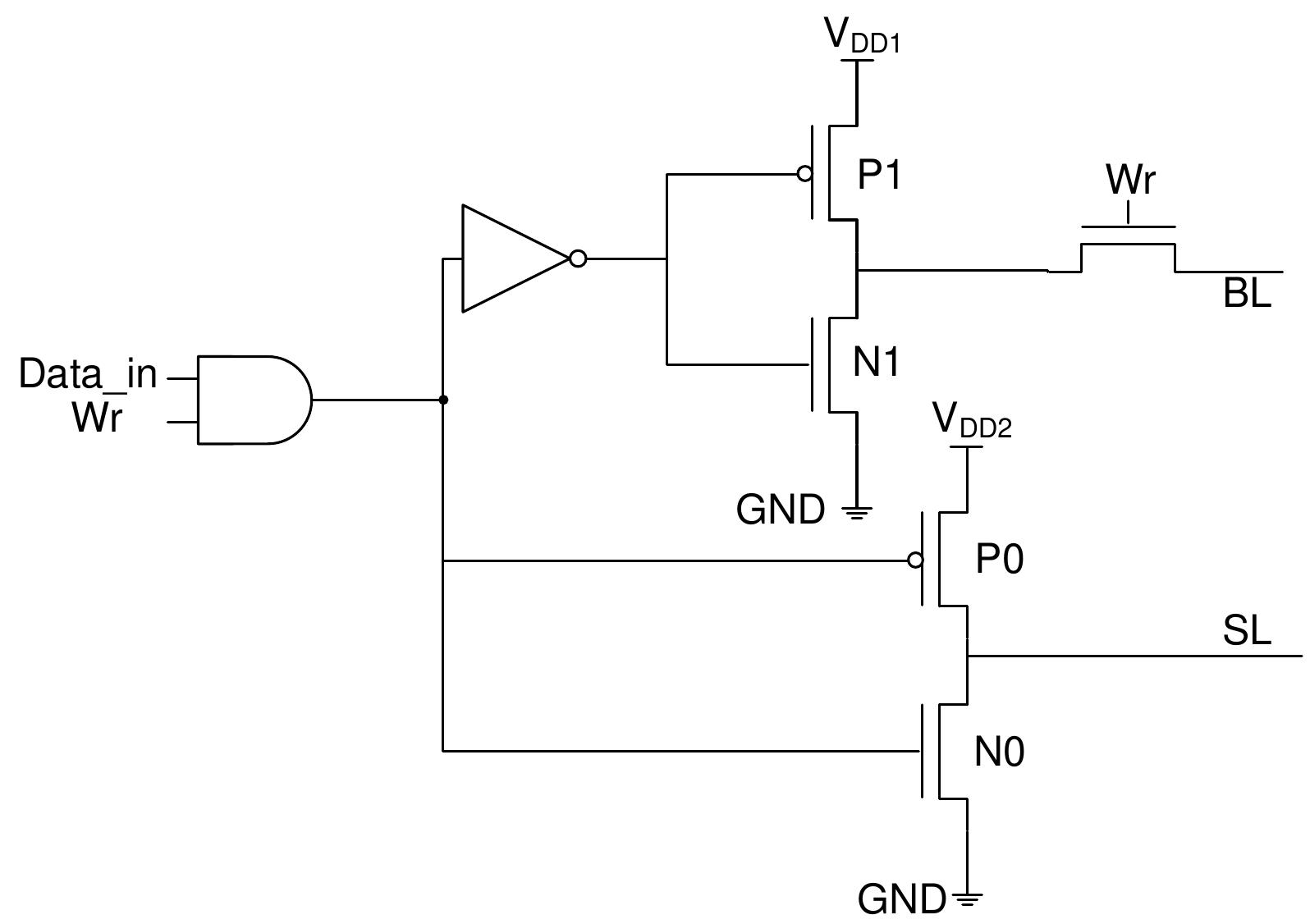}
\caption{Write driver circuit design for RHS-TRNG.} 
\label{fig8}
\end{figure}

The write driver circuit  is shown in Fig.~\ref{fig8}. When the write control signal ``Wr" is enabled and the write data on ``Data\_in'' is at 1, N0 and P1 will be turned on while P0 and N1 will be turned off. This leads to a write current flowing from the BL to the SL. The amplitude and duration of the current determine whether the MTJ will be set to state 1 successfully, as explained in Sec.~\ref{Subsec:STTswitching_stochasticity}.
Likewise, when `Data\_in'' is at 0, a write current with the opposite direction is drawn by the circuit to flip the MTJ's state to 0 with certain probability if it is in state 1.
Since the MTJ device has different resistance values in the P and AP states, the required bias voltages $V_{\mathrm{DD1}}$ and $V_{\mathrm{DD2}}$ are set different to make the currents in the forward and reverse directions have a 50\% switching probability under the same pulse width.

\begin{figure}[t]
\centering
\includegraphics[width=5cm]{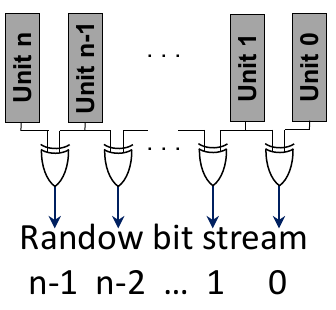}
\caption{ Multi-cell parallel structure of RHS-TRNG.} 
\label{figparall}
\end{figure}

Finally, as illustrated in Fig.~\ref{figparall}, we configure the RHS-TRNG cells in parallel to increase the random bit throughput. To reduce area and power consumption, adjacent cells share an RHS-SingleUnit. The corresponding area and power results are presented in Section.~\ref{sec5}.

\begin{figure}[b]
\centering
\includegraphics[width=8.5cm]{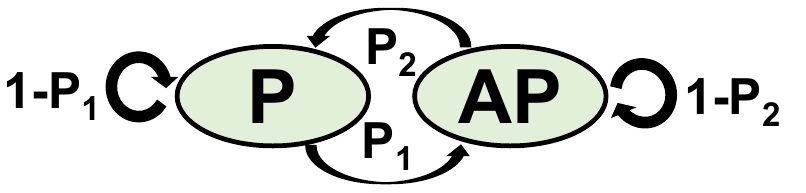}
\caption{State transition diagram of an MTJ device controlled by bidirectional switching currents.} 
\label{fig9}
\end{figure}

\vspace*{-0.2\baselineskip} 
\subsection{Theoretical Analysis}
\label{sub:TheoreticalAnalysis}
\subsubsection{Benefits of Bidirectional Switching Currents}

Compared to the MTJ-based TRNG circuit design with three phases per cycle, our two-phase RHS-TRNG provides lower latency of  random bit generation. Eliminating the reset phase also prolongs the MTJ's lifetime, since the large reset current to ensure a 100\% switching probability is avoided; note that the larger the switching current, the smaller the endurance of the MTJ device. In addition to the above two benefits, our circuit design  provides a resilient mechanism to cope with PVT variations to enhance the quality of random number generation.

\begin{figure}[t]
\centering
\includegraphics[width=8.5cm]{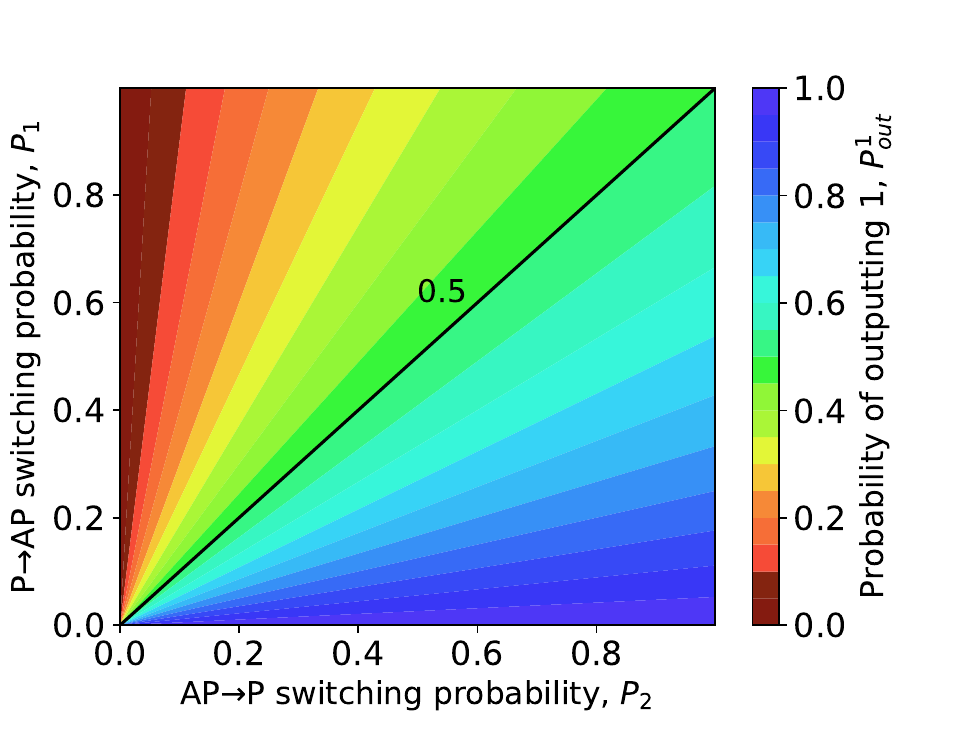}
\caption{The impact of AP$\rightarrow$P and P$\rightarrow$AP switching probabilities on the probability of a random number generator unit outputting logic 1.} 
\label{fig10}
\end{figure}

Fig.~\ref{fig9} shows the MTJ state transition diagram of the TRNG with bidirectional write currents. Initially, the MTJ devices may be in the P state (with the probability $P_{\mathrm{P}}$) or AP state (with the probability $P_{\mathrm{AP}}$). When it is in the P state, we assume that the  probability of being switched to the AP state under a write current  is $P_{1}$. When in the AP state, the  probability is $P_{2}$ under a  write current with the opposite direction. Given these assumptions, we can obtain the following probability relationships in the state transition diagram: 
\begin{equation}
\left\{\begin{array}{l} P_{AP}=P_{AP}\cdot (1-P_{2})+P_{P}\cdot P_{1},
\\P_{AP}+P_{P}=1.
\end{array}\right.
\label{equ1}
\end{equation}
Assume that each generator unit (see Fig.~\ref{fig:overall_circuit_design}) has a probability $P_{\mathrm{out}}^{1}$ of outputting bit 1 and a probability $P_{\mathrm{out}}^{0}$ of outputting bit 0. $P_{\mathrm{out}}^{1}$ and $P_{\mathrm{out}}^{0}$ can be expressed as follows:
%It can be further derived as Equation.~\ref{equ2}:
\begin{equation}
    \left\{\begin{array}{l} P_{\mathrm{out}}^{1}=P_{\mathrm{AP}} =\frac{P_{1}}{P_{1}+P_{2}}, \\
P_{\mathrm{out}}^{0}=P_{\mathrm{P}}=\frac{P_{2}}{P_{1}+P_{2}} .
\end{array}\right.
\label{equ2}
\end{equation}
It can be seen that both  $P_{\mathrm{out}}^{0}$ and $P_{\mathrm{out}}^{1}$depend on two probability variables rather than a single variable as found in some previous designs.

Fig.~\ref{fig10} plots $P_{\mathrm{out}}^{1}$ in Equation~(\ref{equ2}), where the X-axis represents  $P_{2}$ and the Y-axis represents  $P_{1}$. When both $P_{1}$ and $P_{2}$ are equal to 50\% ideally, the probability of the random number generator unit outputting ``1" is also 50\%. The off-center condition is caused by PVT variation. When $P_{1}$ and $P_{2}$ are shifted in the same direction by the same magnitude, $P_{\mathrm{out}}^{1}$  can remain 50\%, as shown with the black solid line in the figure. Recall that several variation sources (e.g., ambient temperature) that may affect the output probability  tend to shift $P_{1}$ and $P_{2}$ along the same direction. Our circuit design allows the TRNG to remain resilient to such variations.

\begin{figure}[t]
\centering
\includegraphics[width=8.8cm]{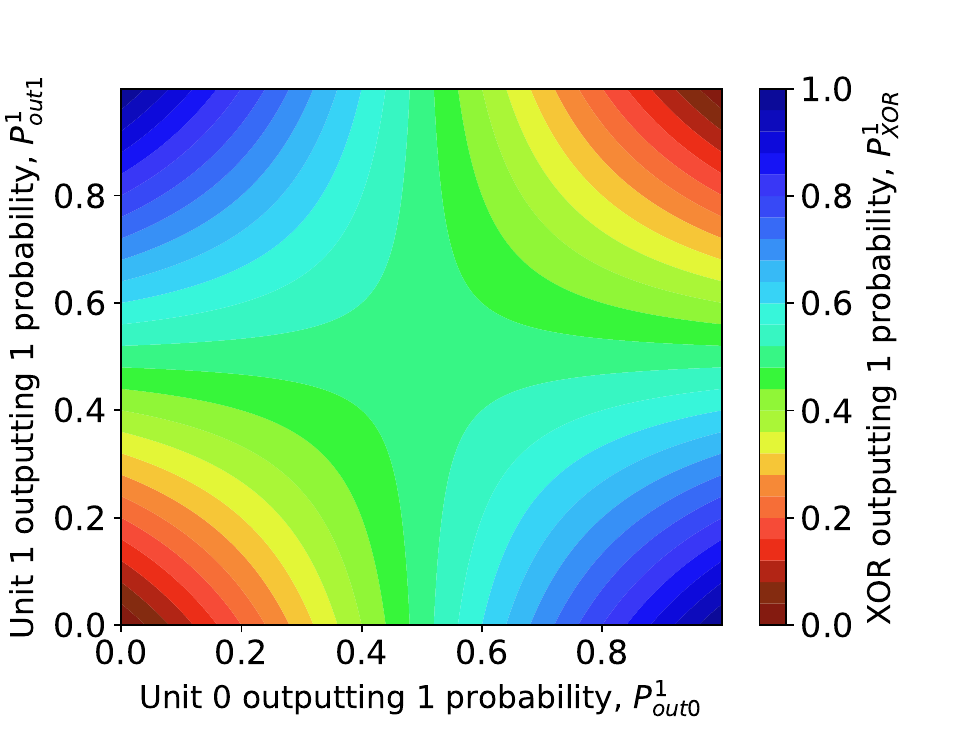}
\caption{Probability of outputting ``1" after XOR of dual generator units.} 
\label{fig11}
\end{figure}

\subsubsection{Benefits of Two Generator Units}

RHS-TRNG uses two identical random number generator units to generate a random bit stream. By means of  this redundant mechanism, the quality of random bit generation is further improved by circuit self-stability. Assume that the probability of the generator unit 0 outputting bit 1 is $P_{\mathrm{out0}}^{1}$ and the probability of the generator unit 1 outputting bit 1 is $P_{\mathrm{out1}}^{1}$, we can mitigate the influence of the  shifts in $P_{\mathrm{out0}}^{1}$ and $P_{\mathrm{out1}}^{1}$ to a certain extent via adding an XOR operation on the outputs of the two generator units. %Through the conditional probability operation, the output probability after XOR of two units is given by Equation.~\ref{equ3}: 
When one of them outputs 1 and the other outputs 0, the XOR result is 1; otherwise, the result is 0.  Thus, the probabilities of the XOR gate  outputting bit 1 ($P_{\mathrm{XOR}}^{1}$) and bit 0 ($P_{\mathrm{XOR}}^{0}$) are calculated as follows:
\begin{equation}
    \left\{\begin{array}{l} P_{\mathrm{XOR}}^{1} =P_{\mathrm{out0}}^{1}\cdot (1-P_{\mathrm{out1}}^{1})+P_{\mathrm{out1}}^{1}\cdot(1-P_{\mathrm{out0}}^{1}), \\
    P_{\mathrm{XOR}}^{0} =1-P_{\mathrm{XOR}}^{1}.
\end{array}\right.
\label{equ3}
\end{equation}
%The probability that a ``1" will be output is shown in Fig.~\ref{fig11}.

Fig.~\ref{fig11} plots $P_{\mathrm{XOR}}^{1}$ as a function of $P_{\mathrm{XOR}}^{0}$ and $P_{\mathrm{XOR}}^{1}$.
It can be seen that when $P_{\mathrm{out0}}^{1}$ and $P_{\mathrm{out1}}^{1}$ are both 50\% ideally, the output result of the entire TRNG design  remains 50\%. If these two values  slightly drift away from  50\%, the output probability $P_{\mathrm{XOR}}^{1}$  still converges to 50\% thanks to the  XOR mechanism.
This design can greatly mitigate the adverse impact of PVT variations, thus keeping the final generated random bit stream stable.
\begin{figure}[b]
	\centering
	\includegraphics[width=7.3cm]{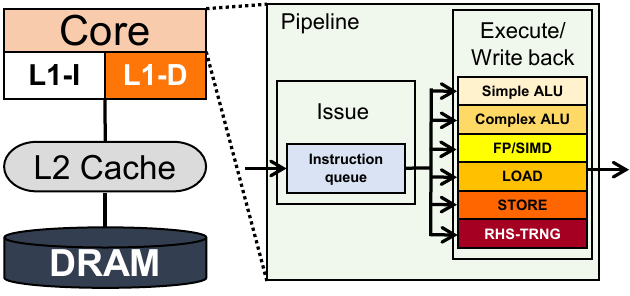}
	\caption{A system overview of multi-core RISC-V processor with the RHS-TRNG hardware integrated into its pipeline as an instruction execution unit.}
	\label{figsystem}
\end{figure}

\begin{figure}[t]
\centering
\includegraphics[width=8cm]{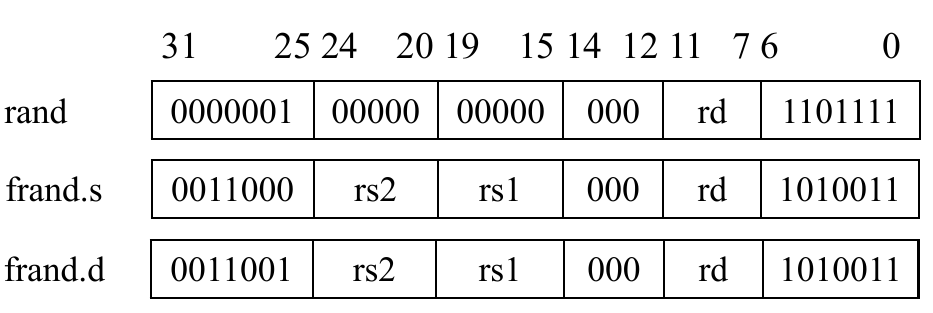}
\caption{The structure of custom random number generation instructions.} 
\label{figinst}
\end{figure}

\begin{figure}[t]
\centering
\includegraphics[width=8cm]{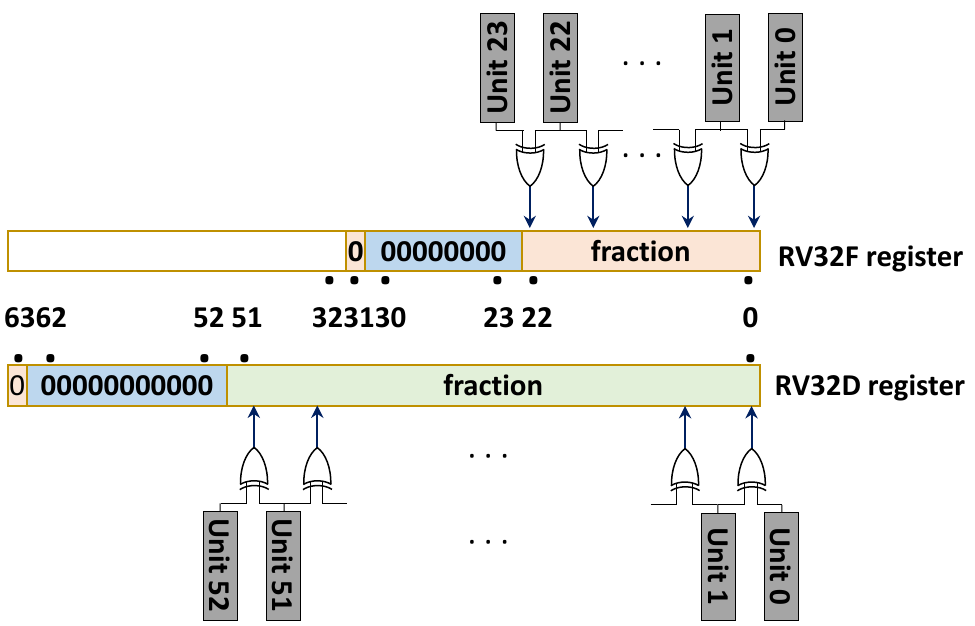}
\caption{Hardware structure for converting integer random numbers into floating-point random numbers.}
\label{figfloattrng}
\end{figure}

\vspace*{-0.2\baselineskip} 
\subsection{System and Custom Instruction Design}
\label{subsec_model}

To unleash the power of our RHS-TRNG hardware in accelerating practical applications, we embed it to the CPU pipeline as an instruction execution unit. Dedicated instructions thus have to be designed to control its execution. Since RISC-V is an open, royalty-free ISA which features design freedom and flexible architecture extensions, our system design is focused on RISC-V processors.

Fig.~\ref{figsystem} shows the architecture of a multi-core RISC-V processor which has integrated our RHS-TRNG as an acceleration component.  The left side of the figure is a general-purpose  four-core architecture. Each RISC-V core owns a private L1 instruction cache (L1-I) and a private L1 data cache (L1-D). All cores share a large-volume L2 cache, which is connected to the main memory (DRAM). On the right side of the figure, we can see  a portion of the instruction pipeline in the core. Note that only the issue and execute stages are shown since they are modified and are directly relevant  to the proposed RHS-TRNG design.

In contrast to a generic RISC-V core, our architecture places an RHS-TRNG as an instruction execution unit in the execution stage. We design a set of custom instruction to control the RHS-TRNG circuit and read the generated random numbers  into a register for use by the program. It should be noted that the integration level of RHS-TRNG into systems depends on many factors such as target application, usage frequency, and lifespan, due to the limited write endurance ($\sim10^{15}$). Leveraging reliability mitigation schemes such as voltage-reducing and wear-leveling can lead to longer lifetime. But this comes with the cost of longer latency and larger area. Thus,  a trade-off between performance and cost has to be made when designing MTJ-based TRNGs. In this paper, we aim to avoid limiting the interface rate for our TRNG. Placing it within the pipeline reduces its read latency and better showcases its performance characteristics.

Fig.~\ref{figinst} shows the design of three custom instructions: rand, frand.s, and frand.d to control the RHS-TRNG acceleration unit. The  instruction ``rand'' controls the generation of an unsigned integer random number between 0 and 32767, it does not require any operand and the generated result is stored in the segment ``rd''. The other two custom instructions ``frand.s'' and  ``frand.d'' control the generation of single and double-precision floating-point numbers,which are converted from integer random numbers and are expanded to the range between the two values stored in rs2 and rs1, respectively. The execution latency of each instruction  depends on the random bit generation period at circuit level as well as the system clock cycle.

For integer random numbers, the RHS-TRNG illustrated in Fig.~\ref{figparall} can directly output parallel bit streams to RV32I registers. However, generating floating-point random numbers requires more considerations. Fig.~\ref{figfloattrng} depicts the method of converting integer random numbers into  single and double-precision floating-point random numbers. Since the exponent bits should not be uniform, we set all the sign and exponent bits to 0 and actually generate non-normalized floating-point numbers ranging from 0 to 1. 

\vspace*{-0.8\baselineskip} 
\section{Experiments and Evaluation}
\label{sec5}

\vspace*{-0.2\baselineskip} 
\subsection{Experimental Method and Setup}

To evaluate the functionality of the proposed self-stabilized RHS-TRNG circuits in this paper, we conducted a series of SPICE circuit simulations. The Verilog-A MTJ compact model in \cite{wang2014compact} is used in our simulations; it  models the TMR effect and STT effect of the MTJ. As a physical model that integrates static, dynamic, and stochastic behaviors, it is currently widely used by researchers in the field of Spintronics. The MTJ device parameters  are shown in Table.~\ref{tab1}; their values are selected in accord with real-world MTJs fabricated at IMEC. We set the ambient temperature to \SI{300}{K} (i.e., room temperature) except for the study of temperature variation. The proposed design is simulated using the Cadence Virtuoso tool with GPDK \SI{45}{\nano\metre} technology. 

\begin{table}[t]
\centering
\caption{Key device parameters for MTJ compact model.}\label{tab1}
\resizebox{0.5\textwidth}{!}{
\begin{tabular}{
>{}c |
>{}c |
>{}c }
\hline
{ \textbf{Parameter}} & {\textbf{Description}}  & {\textbf{Value}}  \\ \hline \hline
{$t_{\mathrm{FL}}$}       & {Thickness of the free layer}        & {1.3nm}  \\ \hline
{$\sigma _{t_{\mathrm{FL}}} $}       & {Standard deviation of $t_{\mathrm{FL}}$}        & {3\% of 1.3nm}  \\ \hline

{$CD$}         & {Critical diameter}        & {32nm}   \\ \hline
{$t_{\mathrm{TB}}$}       & {Thickness of the tunnel barrier}     & {0.85nm} \\ \hline

{$\sigma _{t_{\mathrm{TB}}} $}       & {Standard deviation of $t_{\mathrm{TB}}$}     & {3\% of 0.85nm} \\ \hline

{$TMR$}       & {TMR ratio} & {200\%}   \\ \hline
{$\sigma _{TMR}$}       & {Standard deviation of TMR} & {3\% of 200\%}   \\ \hline
\end{tabular}
}
\end{table}

\begin{table}[b]
\centering
\caption{Simulated system configurations  for the gem5 simulator.}\label{tab2}
\resizebox{0.4\textwidth}{!}{
\begin{tabular}{c|c}
%\hline
\textbf{Parameter}         & \textbf{Value}    \\ \hline \hline
Gem5 version      & 21.1.0   \\ \hline
Simulation model  & SE       \\ \hline
CPU Type          & MinorCPU \\ \hline
Frequency     & \SI{2}{GHz}        \\ \hline
Icache size   & \SI{32}{kB}        \\ \hline
Dcache size    & \SI{32}{kB}      \\ \hline
L2 cache size  & \SI{512}{kB}       \\ \hline
Instruction execution latency     & 8  cycles      \\ \hline
\end{tabular}
}
\end{table}

In our experiments, we first evaluated the timing performance of RHS-TRNG  to determine the maximum throughput. Then we obtained a random sequence with a length of 1 million via Monte Carlo simulation to evaluate its statistical randomness. Next, we conducted experiments to analyze the quality of generated random sequences under PVT variations.
The area and power consumption of the proposed circuit were then evaluated by the simulation results and layout design results using the Cadence Virtuoso tool. To systematically evaluate the acceleration effect of RHS-TRNG in a general-purpose computing system for the corresponding application, we also integrated it into a RISC-V processor as an instruction execution unit and modeled the system in the gem5 simulator. 
The architecture that incorporates this unit has significant performance advantages over the general-purpose architecture when running a Monte Carlo option pricing program. 
The architectures used for comparison are  identically configured, as shown in Table.~\ref{tab2}.

To compare the performance of different MTJ-based TRNGs including both the conventional three-stage design and our two-stage design in this work, we implemented four  TRNG configurations as follows.

\begin{enumerate}
\item \textbf{Conv. APtoP}:
\\
This configuration is a conventional (Conv.) MTJ-based TRNG design (see Fig. \ref{figmotiv}), where  the WRITE phase applies an APtoP switching current flowing through the MTJ device to generate a random bit.

\item \textbf{Conv. PtoAP}:
\\
This configuration is identical to the above Conv.APtoP configration except that the current in the WRITE phase has a probability of  50\% to switch the MTJ  from the P state to the AP state.

\item \textbf{RHS-SingleUnit}:
\\
This configuration is a single random number generator unit of RHS-TRNG (i.e., Unit 0 or Unit 1 in Fig.~\ref{fig:overall_circuit_design}). Its MTJ cell and peripheral circuits are consistent with that of RHS-TRNG. 

\item \textbf{RHS-TRNG}:
\\
This configuration represents a complete RHS-TRNG design.
\end{enumerate}

The names of these configurations will be used directly in the remainder of this paper.

\vspace*{-0.2\baselineskip} 
\subsection{SPICE Circuit Simulation}
\begin{figure}[t]
\centering
\includegraphics[width=8cm]{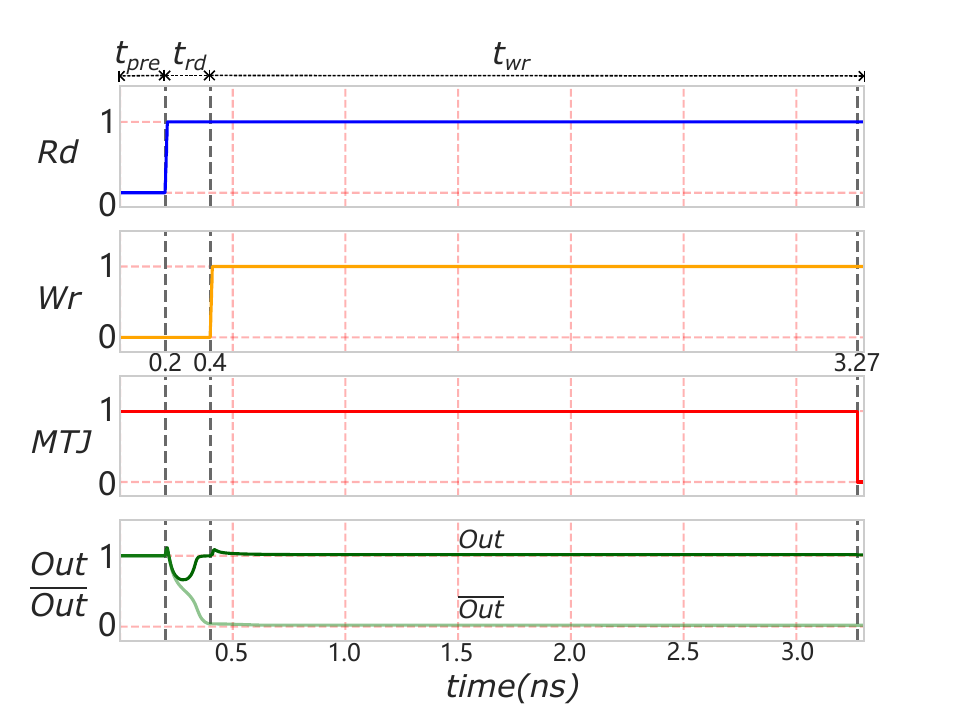}
\caption{Waveforms of key signals in a transient circuit simulation of RHS-SingleUnit for a single random bit generation cycle (\SI{3.3}{\nano\second}).} 
\label{figlogictime}
\end{figure}

As our RHS-TRNG design consists of two identical RHS-SingleUnits and an XOR gate, we ran transient simulations using the RHS-SingleUnit configuration for the sake of similarity in the Cadence virtuoso tool to evaluate its performance.  Fig.~\ref{figlogictime} shows the transient  simulation results, i.e., the waveforms for  the ``Rd" signal, the ``Wr" signal, the state of the MTJ, and the readout voltage at node $\mathrm{Out}$ and $\overline {\mathrm{Out}}$ in Fig.~\ref{fig7}. Recall that each cycle contains two phases: read and write; the read phase is further divided into three sub-phases: pre-charge, voltage development, and voltage amplification. 

First, in the read phase, the ``Wr" and ``Rd" signals are both set to low potential. After the sense amplifier completes the pre-charge sub-phase (as short as $t_{\mathrm{pre}}$), the ``Rd" signal is pulled high. This is followed by the voltage development and amplification sub-phases; the time periods are together denoted as $t_{\mathrm{rd}}$ in the figure).  After the read phase, the current MTJ state is read out at the node $\mathrm{Out}$.
Next is the write phase, where the ``Wr" signal is pulled high to write the opposite state to the MTJ with a 50\% probability. In the case shown in the figure, the MTJ switches within $t_{\mathrm{wr}}$. Because of the probabilistic switching behavior, we can achieve random bit generation by setting $t_{\mathrm{wr}}$ to the mean of the random switching time distribution. 

Through extensive experiments, we observed that $t_{\mathrm{pre}}$ and $t_{\mathrm{rd}}$ in Fig.~\ref{figlogictime} can be controlled below \SI{0.2}{\nano\second}, respectively, and $t_{\mathrm{wr}}$ can be controlled at about \SI{2.9}{\nano\second}. Hence, the random bit generation latency of each RHS-SingleUnit can be controlled at \SI{3.3}{\nano\second/bit}. In summary, RHS-TRNG can achieve a generation rate of \SI{303}{Mb/s} for a single cell, and will provide higher random bit output rates when parallelized as needed.

\vspace*{-0.2\baselineskip} 
\subsection{Statistical Randomness Test}

The National Institute of Standards and Technology (NIST) SP 800–22 rev.1a Test\cite{bassham2010sp} is used to test whether a set of binary bit sequences satisfies statistical randomness; it is often used to test the random number generation quality of RNGs. The NIST test consists of different  test modules that examine the  presence of non-random samples in a sequence of bits. For example, the  test module ``Frequency'' checks whether the appearance frequencies of ``0" and ``1" in a bit sequence are approximately the same. The test module ``Runs'' is used to test whether the number of consecutive occurrences of the same bit such as ``0000" or ``111" is as expected. ``FFT'' is used to detect the peak height after the step-by-step discrete Fourier transform of a sequence, thus detecting the periodicity of the signals under test. Each test gives  a  P-value which quantifies the difference between a  sample  bit sequence under test and an ideal random bit sequence. When the P-value is greater than the threshold, the sample can be considered to pass this test. If the sequence length is large enough, some of the tests are executed multiple times and further give a pass rate. We consider the sample to pass the test when both the P-value and the pass rate are greater than their respective thresholds. Generally, the P-value threshold is 0.0001, and the pass rate threshold is 0.91. For a given  bit sequence which passes all tests, it is considered to have good distribution characteristics in terms of statistical randomness.

We generated one million random bits for NIST tests and divided them into 10 groups. To generate random bit sequences that avoid artificially inducing favorable results, we simulated the switching of MTJs from AP to P state and P to AP state separately, and then mixed them according to the mechanism proposed in our design, which captures the correlation between periods. The test results are shown in Table.~\ref{tab3}. It can be seen that we passed all the test modules in the table, which suggests that  bit  streams generated by RHS-TRNG have excellent statistical randomness.

\begin{table}[t]

\caption{NIST test results  on random bit sequences generated by RHS-TRNG.}
\label{tab3}
\vspace{-20pt}
\begin{center}
\resizebox{0.5\textwidth}{!}{
			\bgroup
			\def\arraystretch{1.2}%  1 is the default, change whatever you need
\begin{tabular}{|cc|c|c|c}

\multicolumn{2}{c|}{\textbf{Test module}}  & \textbf{P-value} & 	\textbf{Pass rate} & \textbf{Pass/Fail}  \\ \hline \hline
\multicolumn{2}{c|}{Frequency}                                 & 0.911413 & 10/10     & Pass      \\ \hline
\multicolumn{2}{c|}{BlockFrequency}                            & 0.911413 & 10/10     & Pass      \\ \hline
\multicolumn{1}{c|}{CumulativeSums} & Forward & 0.213309 & 10/10     & Pass      \\ \cline{2-5} 
\multicolumn{1}{c|}{}                                & Reverse & 0.350485 & 10/10     & Pass      \\ \hline
\multicolumn{2}{c|}{Runs}                                      & 0.739918 & 10/10     & Pass      \\ \hline
\multicolumn{2}{c|}{LongestRun}                                & 0.350485 & 10/10     & Pass      \\ \hline
\multicolumn{2}{c|}{Rank}                                      & 0.350485 & 10/10     & Pass      \\ \hline
\multicolumn{2}{c|}{FFT}                                       & 0.991468 & 10/10     & Pass      \\ \hline
\multicolumn{2}{c|}{NonOverlappingTemplate}                    & -        & 1460/1480 & Pass      \\ \hline
\multicolumn{2}{c|}{OverlappingTemplate}                       & 0.534146 & 10/10     & Pass      \\ \hline
\multicolumn{2}{c|}{ApproximateEntropy}                        & 0.213309 & 10/10     & Pass      \\ \hline
\multicolumn{1}{c|}{Serial}         & Forward & 0.213309 & 9/10      & Pass      \\ \cline{2-5} 
\multicolumn{1}{c|}{}                                & Reverse & 0.122325 & 9/10      & Pass      \\ \hline
\multicolumn{2}{c|}{LinearComplexity}                          & 0.739918 & 10/10     & Pass      \\ \hline

\end{tabular} 
			\egroup
		}
	\end{center}%
%	\vspace{-10pt}%
\end{table}

\vspace*{-0.2\baselineskip} 
\subsection{Resilience Experiments Against PVT Variations}

To evaluate the resilience enhancement effect of the two self-stabilization mechanisms in RHS-TRNG introduced in Section.~\ref{sec4}, we experimentally evaluated its ability of tolerating PVT variations. The experimental results are also compared to that of traditional three-stage MTJ-based TRNG designs. 

\subsubsection{Output Entropy Concept}
In our experiments, we use output entropy to quantify the random bit generation quality of TRNG. Entropy describes the chaotic degree of information in a system. The higher the entropy value, the more chaotic the system  is. For a bit sequence, a high entropy value indicates that the sequence is  well uniformly distributed. 

Typically, two types of entropy are widely used, which are Shannon entropy and minimum entropy.
For each bit in a sequence, its state is either ``0" or ``1".
Assume that each bit  is a random variable $X$, the Shannon entropy of this sequence is defined as:
\begin{equation}
    H_{Shannon}(X)=-\sum_{x}^{}  P(x)\log_{2}{P(x)},  x\in\{0, 1\},
\label{equ4}
\end{equation}
where $P(x)$ is the occurrence probability of $x$ in the bit sequence.
The range of $ H_{Shannon}(X)$ is  $[0,\log_{2}{m} ]$, where $m$ is the number of all possible states of $X$ ($m=2$ in this case).
One can easily derive that the maximum Shannon entropy  is 1 when the distribution probabilities of ``0'' and ``1'' are both 1/2. Similarly, for such a random variable X, its minimum entropy is defined as:
\begin{equation}
    H_{Min}(X)=min(-\log_{2}{P(x)} ), x\in\{0, 1\}.
\label{equ5}
\end{equation}
$H_{Min}(X)$ is also in the range $[0, 1]$. When the distribution of the bit sequence is non-random, the Shannon entropy is less than 1 and the minimum entropy is even smaller. The minimum entropy is the lower bound of  entropy and represents the worst distribution of the random variable reflected by a sample. Combined with the Shannon entropy, we can estimate the range of fluctuation in the randomness of the distribution of the random variable.

\begin{figure}[t]
\centering
\includegraphics[width=8cm]{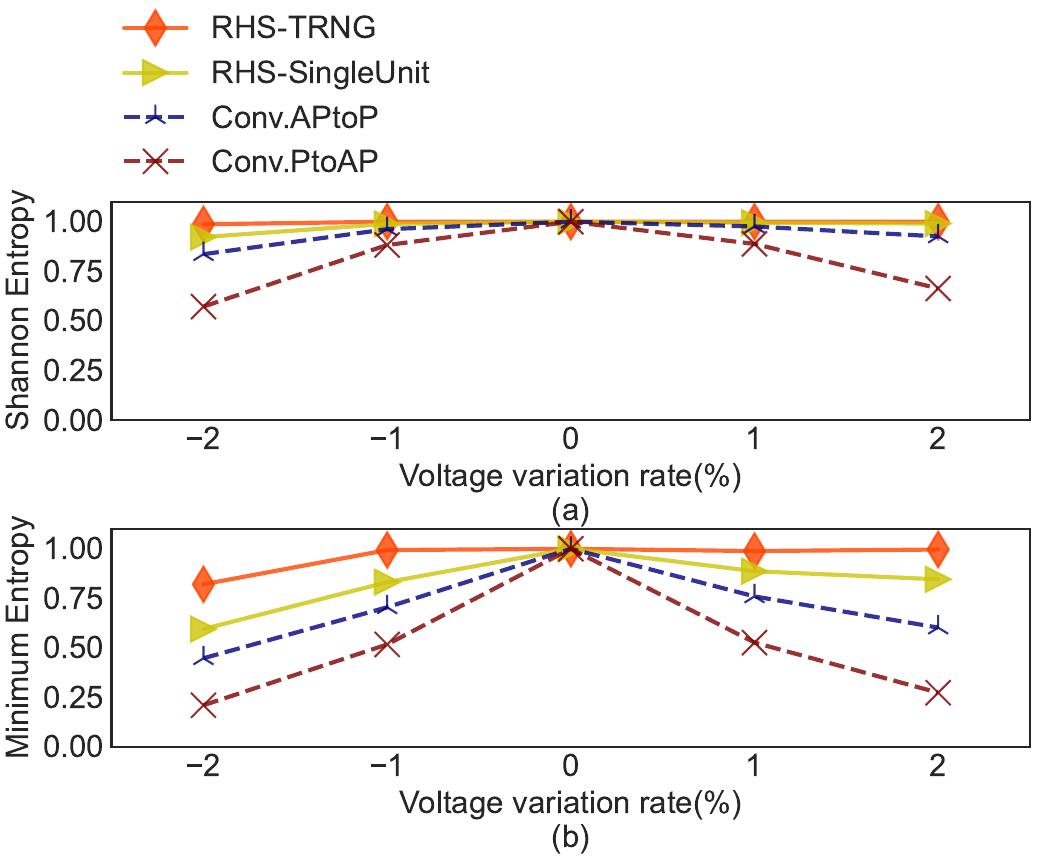}
\caption{Entropy changes of the generated random sequences using different TRNG designs under voltage variation.} 
\label{figvv}
\end{figure}

\begin{table}[b]
	\caption{Output entropy of different TRNG designs when parameter variation of MTJ equipment is considered.}
    \label{tabadd}
	\vspace{-20pt}
	\begin{center}
		\resizebox{0.5\textwidth}{!}{
			\bgroup
			\def\arraystretch{1.2}%  1 is the default, change whatever you need
			\begin{tabular}{c|c|c|c}		
           & \textbf{RHS-TRNG} & 	\textbf{Conv.APtoP} & \textbf{Conv.PtoAP}  \\ \hline \hline
       Shannon entropy &  0.99996 & 0.94742   &0.86682   \\ \hline 
         Minimum entropy & 0.99001  & 0.65707     & 0.49113  \\\hline 
           
			\end{tabular} 
			\egroup
		}
	\end{center}%
%	\vspace{-10pt}%
\end{table}

\subsubsection{Resilience Against Voltage Variation}

In Section.~\ref{sub:TheoreticalAnalysis}, we have theoretically evaluated the improved output probabilistic stability using bidirectional switching currents and two generator units.
We also performed solid experiments to evaluate the resilience advantages brought by the above-mentioned two mechanisms when compared to traditional TRNG designs. We varied the supply voltages $V_{\mathrm{DD1}}$ and $V_{\mathrm{DD2}}$ in a stepped manner, as shown in Fig.~\ref{fig8}, for four configurations of MTJ-based TRNGs: Conv.APtoP, Conv.PtoAP, RHS-SingleUnit, RHS-TRNG. We calculated the entropy values for the generated random sequences under voltage variation; the results are shown in Fig.~\ref{figvv}.a and Fig.~\ref{figvv}.b for Shannon entropy and minimum entropy, respectively. When the voltage variation rate is 0, it means that  $V_{\mathrm{DD1}}$ and $V_{\mathrm{DD2}}$ are at nominal values. A positive voltage variation rate presents an increase in the two supply voltages, while a negative value indicates a decrease in the voltages. The y-axis in the figure represents the entropy value in response to voltage variation; the closer to 1, the higher quality of the random sequence. 

It can be seen that RHS-SingleUnit has higher Shannon entropy and minimum entropy than the two conventional TRNG designs, indicating that our design using bidirectional switching currents can increase the resilience of the circuit to voltage variation. We can also see that the Shannon entropy and minimum entropy of the sequence output by RHS-TRNG are significantly higher than those of RHS-SingleUnit, which suggests that the two generator unit mechanism can further improve the resilience of the circuit to voltage variation.

\subsubsection{Resilience Against Process Variation}

Similar to transistors, MTJ device parameters also fluctuate around their nominal values due to process variation. We took into account three key parameters  $t_{\mathrm{FL}}$, $t_{\mathrm{TB}}$, and $TMR$  in our experiments, where we assigned a Gaussian distribution to them with a variation percentage of 3\% according to the MTJ model\cite{wang2014compact}, as shown in Table.~\ref{tab1}. The experimental results are shown in Table.~\ref{tabadd}. It can be seen that under the influence of  process variation, the minimum entropy of the sequence output by RHS-TRNG is still greater than 0.99. However, the minimum entropy of the other two sequences generated by the conventional MTJ-based TRNGs has dropped to 0.49 and 0.66. Due to the fact that process variations in MTJ devices have a significant impact on resistance,  the resultant drop in output entropy is unacceptable without any mitigation designs. With our proposed scheme using bidirectional switching currents and two generator units, RHS-TRNG can effectively tolerate device parameter variations and guarantee high output entropy.

\begin{figure}[t]
\centering
\includegraphics[width=7.5cm]{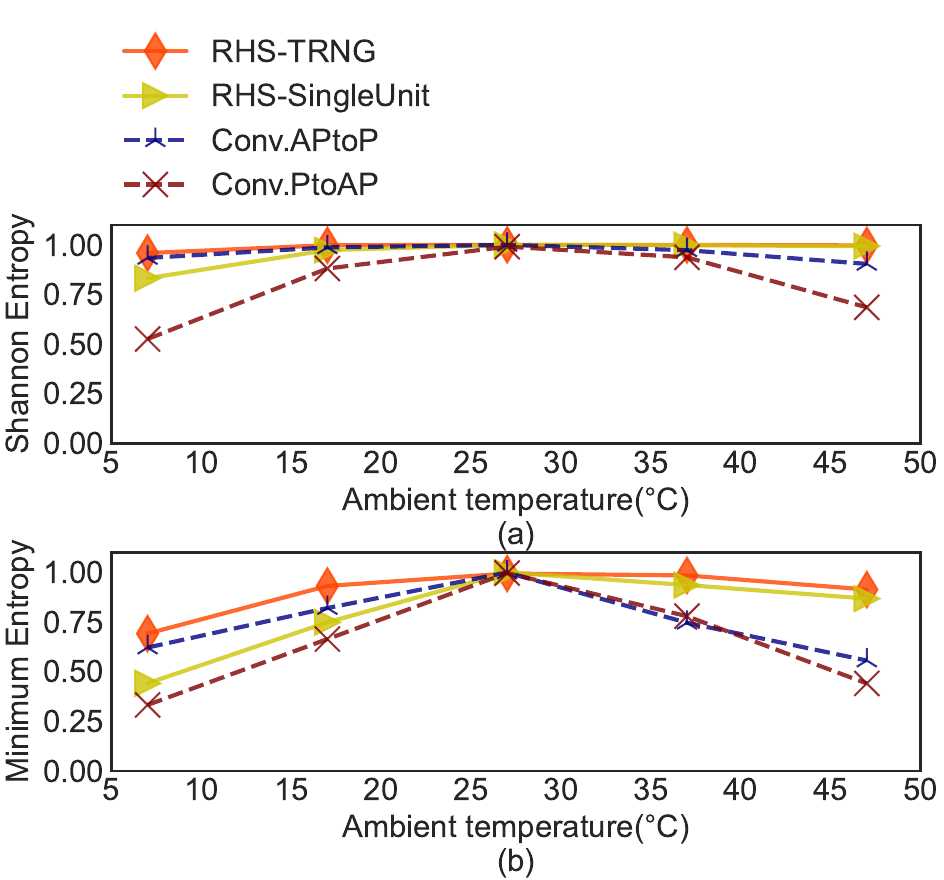}
\caption{Output entropy of different TRNG designs at different temperatures.} 
\label{fig14}
\end{figure}

\subsubsection{Resilience Against Temperature Variation}

MTJ-based TRNG  circuits may also be affected by various external conditions. This paper takes the ambient temperature variation as an example to study the ability of RHS-TRNG in resisting environmental variations. Fig.~\ref{fig14} shows the impact of ambient temperature on Shannon entropy and minimum entropy for the four different configurations.
The  temperature is swept from \SI{7}{\degreeCelsius} to \SI{47}{\degreeCelsius}, and the center of the x-axis is  \SI{27}{\degreeCelsius}. Fig.~\ref{fig14}.a shows that the Shannon entropy of RHS-TRNG is always close to 1 across the entire temperature range. Compared to Conv.APtoP and Conv.PtoAP, the RHS-TRNG  output sequence also provides higher minimum entropy. Interestingly, it can be observed that the ability of RHS-TRNG to withstand high-temperature changes is stronger than its ability to withstand low-temperature changes. These results indicate that the  RHS-TRNG circuit is suitable for being integrated into a computer system as an acceleration component. We can also note that RHS-SingleUnit performs even less well in low temperatures than Conv.APtoP, which indicates that the bidirectional switching current mechanism does not cope well with low temperature variations. But when combined with two generator units mechanism, we can obtain the optimal output entropy.

\begin{figure}[t]
\centering
\includegraphics[width=7cm]{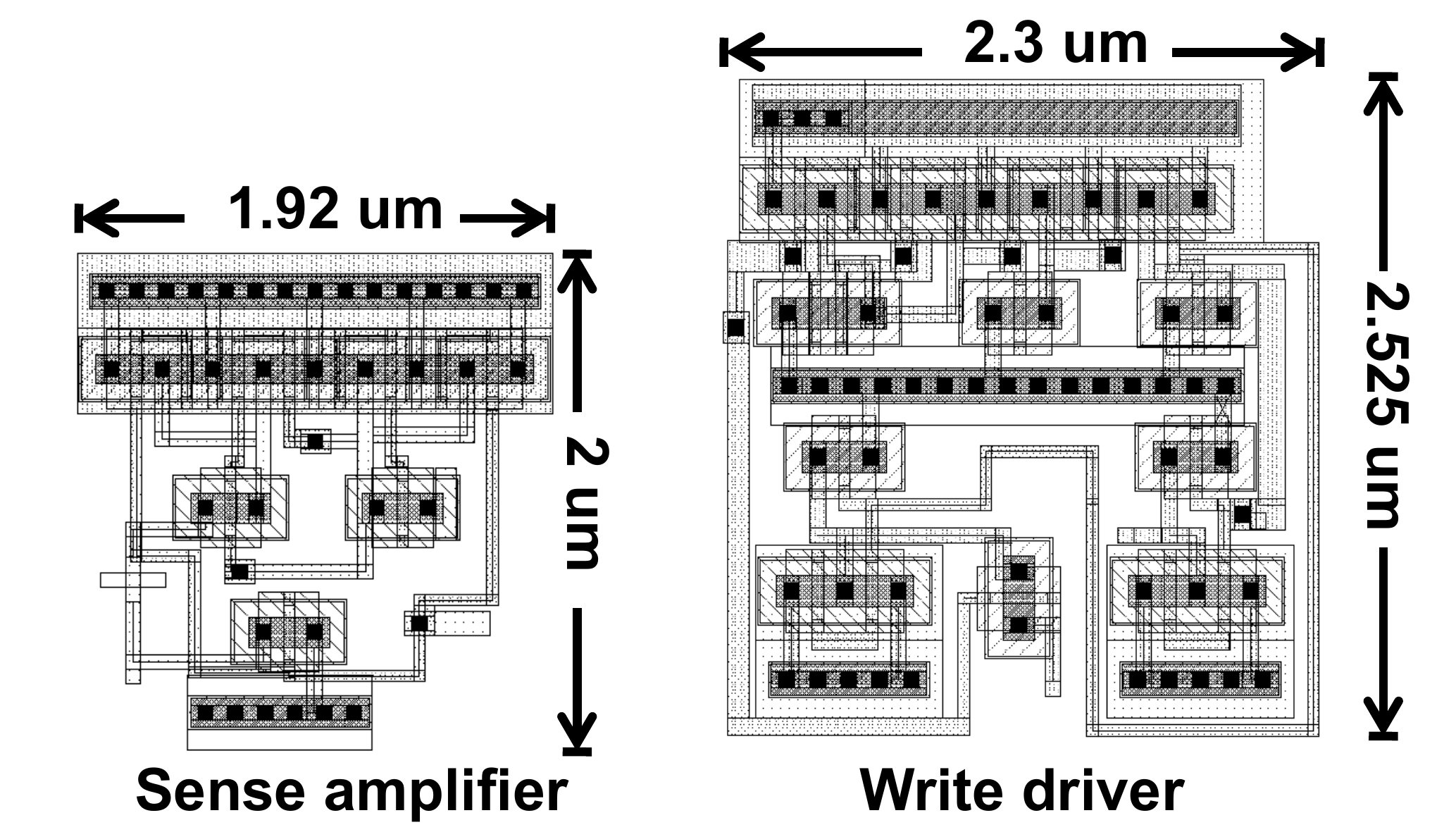}
\caption{Layout of the sense amplifier and the write driver for RHS-TRNG.}
\label{figlayout}
\end{figure}

\begin{figure*}[t]
\centering
\includegraphics[width=18cm]{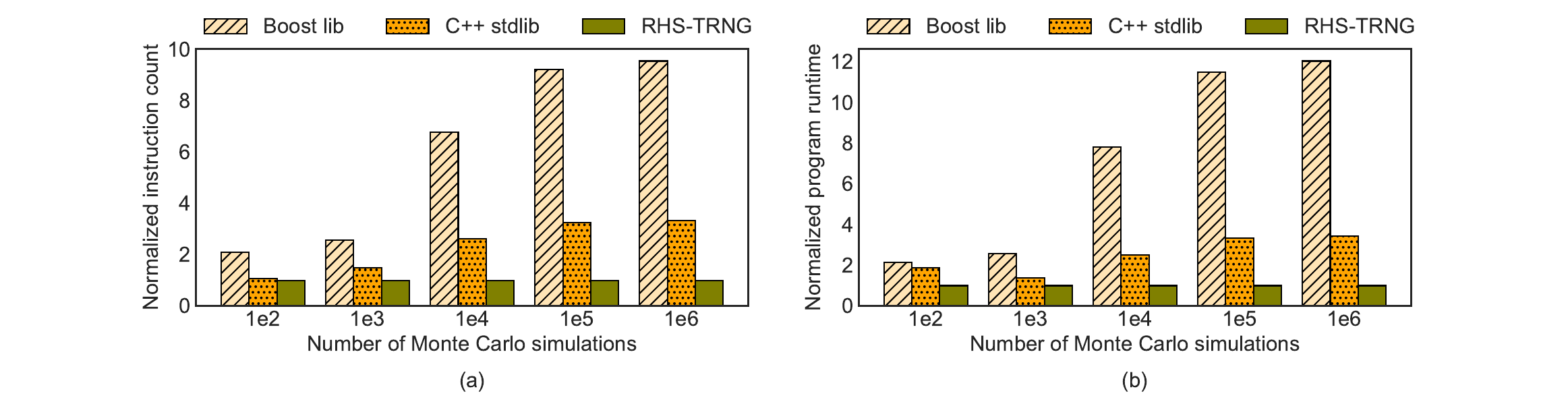}
\caption{System performance comparison using option pricing benchmark which generates random numbers by Boost lib, C++ stdlib, and RHS-TRNG.} 
\label{fig15}
\end{figure*}

\vspace*{-0.2\baselineskip} 
\subsection{Area and Power}

We built RHS-TRNG circuits in  Cadence Virtuoso; the circuit power consumption was evaluated through transient simulation and the on-chip area was evaluated through layout design. Using the GPDK \SI{45}{\nano\metre} technology, the power consumption of RHS-TRNG is \SI{1.6}{mW} for generating a random  bit in a generation cycle of \SI{3.3}{\nano\second}; that is, the power consumption of the TRNG is \SI{5.3}{\pico\joule}/bit. The area of the MTJ device is based on the results presented by Vincent et al.\cite{vincent2014analytical}. Fig.~\ref{figlayout} shows the layout design of the sense amplifier and write driver, with an area of \SI{9.64}{\micro\metre^2} together. To form an RHS-SingleUnit, an NMOS transistor and an MTJ have to be added, resulting in a total area of \SI{9.79}{\micro\metre^2}. Furthermore, to construct a complete RHS-TRNG, two such units and an XOR gate are required, leading to a total area of \SI{24.29}{\micro\metre^2} (see Fig. \ref{fig:overall_circuit_design}).

Our RHS-TRNG design also shows great scalability to achieve higher throughput. After further parallel expansion as shown in Fig.~\ref{figparall}, the n parallel output bit sequences can be genarated by n+1 RHS-SingleUnits, with a single output bit occupying an area slightly larger than \SI{14.5}{\micro\metre^2} and consuming slightly more than \SI{2.65}{\pico\joule}. Although RHS-TRNG does not show outstanding performance in terms of power and area,  the evaluation results are still within an acceptable range. It is worth noting that the power and area  will be greatly reduced using more advanced MTJ and CMOS technologies. Detailed comparisons with other works will be presented in the next section. 

\vspace*{-0.2\baselineskip} 
\subsection{System-Level Evaluation}

By modeling RHS-TRNG in the architecture simulator, we can evaluate the performance acceleration of the benchmark program when it is integrated into the system. 

In the extreme case, one random bit generation cycle of our TRNG unit consists of two processes with times of \SI{0.4}{\nano\second}, and \SI{2.9}{\nano\second}, respectively, so the highest supported main frequency can reach 3.3GHz. We list the dominant frequencies of the architecture modeled in the simulator in Table.~\ref{tab2}. Therefore, to adapt to the system clock, the running time of the three stages of TRNG will be relaxed to \SI{0.5}{\nano\second}, \SI{0.5}{\nano\second} and \SI{3}{\nano\second}, and the delay of each TRNG generation instruction is 8 ticks. We ran the Monte Carlo option pricing program in SE mode using gem5 to evaluate the acceleration effect at the system level.

\subsubsection{Monte Carlo Option Pricing Benchmark}

The Monte Carlo algorithm is widely used in scientific computing, finance, radiology and other fields. It is a representative class of applications that require massive high-quality random numbers. Monte Carlo simulation is often used for option pricing, risk management, and financial modeling in the financial field. It can deal with complex high-dimensional problems that are difficult to solve using traditional analytical methods, but the simulation time has always been a big concern for researchers. Malesevic introduces the background knowledge and corresponding procedures of option pricing using the Monte Carlo method in \cite{malesevic2017use}. The benchmarks used in this paper are derived from the General Monte Carlo Method listed there. We rewrite the program of Hilpisch et al.\cite{hilpisch2014python} into a C++ program that can be compiled and run on gem5.
The benchmark program can be configured using the following three methods to generate random numbers: 
1) the rand function of the C++ stdlib, 2) the lagged\_fibonacci1279 function of the Boost lib\cite{schaling2011boost}, and 3) the proposed custom RHS-TRNG instructions (see \ref{subsec_model}).

\subsubsection{Performance Comparison}

We ran the option pricing benchmark with the above-mentioned three different configurations on gem5. The system configurations can be found in Table.~\ref{tab2}.

Fig.~\ref{fig15}.a compares the instruction count of the benchmark for the three  random number generation methods.  We normalized all values to that of RHS-TRNG for the sake of comparison. When the number of Monte Carlo simulations is 1e2, the instruction counts for the benchmark using Boost lib and C++ stdlib are 2 and 1 times larger than that of the RHS-TRNG, respectively. This is because when the number of simulations is small, the random number generation part does not occupy much of the total program runtime.  
As the Monte Carlo simulation number increases, the advantage of RHS-TRANG-enpowered system starts to stand out. It can be seen that these two multiples will converge at 9.5 and 3.3 times When the simulation time reaches 1e6. This is due to the fact that after reaching the number of simulations that make the results converge, the main overhead of the program is spent on generating random numbers, of which our custom instructions reduce the number of instructions significantly.

Fig.~\ref{fig15}.b illustrates the difference of program runtime of the benchmark at different Monte Carlo simulation numbers. We can observe that when the simulation number exceeds 1e5, the speedup effect of our RHS-TRNG on the option pricing program starts to level off. In summary, using RHS-TRNG achieves 3.4–12$\times$ performance acceleration in comparison to the other two software-based RNGs. This result is consistent with the improvement in the number of instructions.

\begin{figure}[t]
\centering
\includegraphics[width=8.5cm]{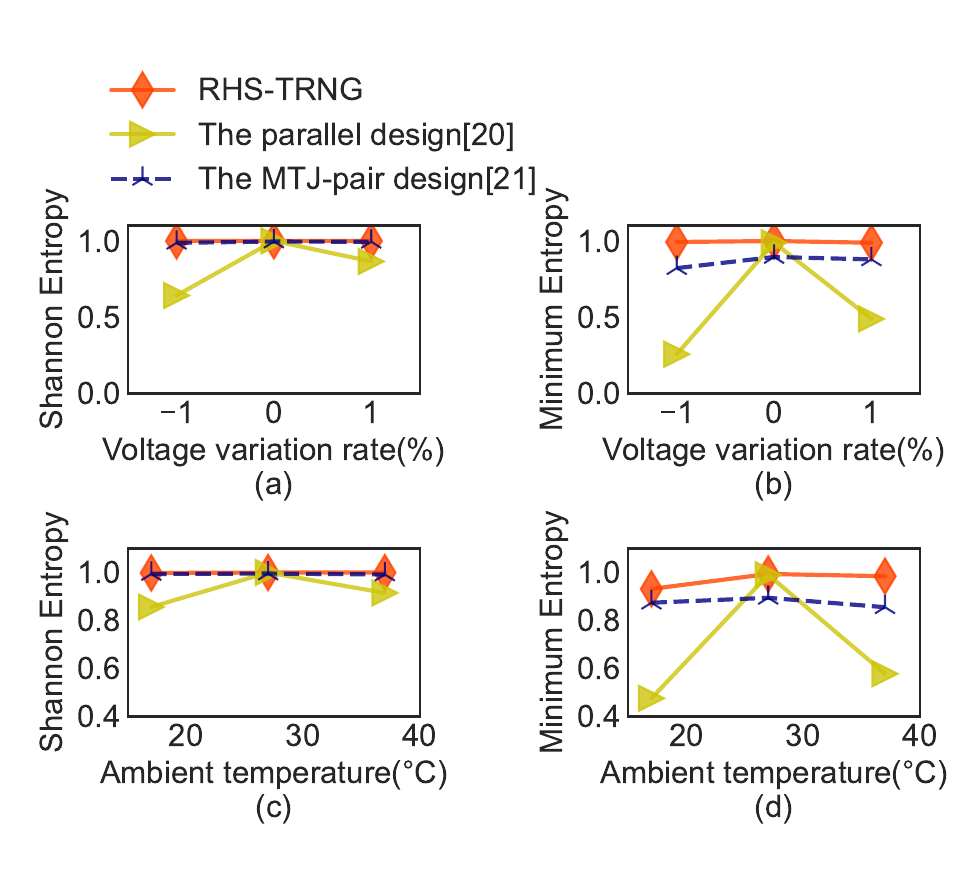}
\caption{Comparison of Shannon entropy and minimum entropy between RHS-TRNG, the parallel design in \cite{qu2017true}, and the MTJ-pair design in \cite{qu2018variation} under voltage and temperature variations.} 
\label{compare}
\end{figure}

\begin{figure}[t]
\centering
\includegraphics[width=8.5cm]{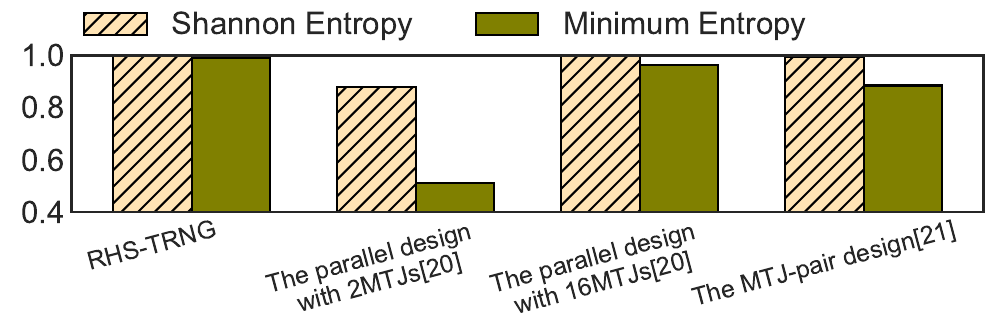}
\caption{Comparison of Shannon entropy and minimum entropy under MTJ process variation between RHS-TRNG, the parallel design in \cite{qu2017true}, and the MTJ-pair design in \cite{qu2018variation}.} 
\label{compare_p}
\end{figure}

\vspace*{-0.8\baselineskip} 
\section{Related Work}
\label{sec6}

In this section, we compare our RHS-TRNG design with some state-of-the-art TRNG designs, including DRAM-based and MTJ-based, and highlight the advantages of this work.

Olgun et al.\cite{olgun2021quac} proposed a quadruple-activation (QUAC) TRNG design based on commodity DRAM chips. QUAC-TRNG activates four memory rows that store conflicting data simultaneously through several consecutive DRAM commands that violate timing constraints. This causes the bitline sense amplifiers to non-deterministically converge to random values, which are considered as a random bit sequence. QUAC-TRNG generates 7664 random bits every \SI{1940}{\nano\second}, equivalent to a throughput rate of \SI{3.44}{Gb/s} per DRAM channel. Despite the claimed high throughput, QUAC-TRNG is based on the configuration of bank group parallelism and row clone, which requires the use of a whole DRAM block; it also conflicts with normal DRAM accesses. In contrast, RHS-TRNG  obtains a tenth of its throughput  with only a single memory cell. By exploiting cell-level parallelism, the throughput of RHS-TRNG can easily outperform that of QUAC-TRNG. Moreover, RHS-TRNG consumes much less energy since it does not require the periodic refresh operations ($\sim\SI{60}{\milli\second}$) in DRAM.

Perach et al.\cite{perach2019asynchronous} designed an MTJ-based asynchronous TRNG for low-power edge devices. By using the capacitor discharge as the current excitation source for the MTJ, this asynchronous design decouples the random number generation process from the system clock. In terms of entropy generation rate, a random bit generation cycle is divided into a charging phase, a enable phase, and a read phase. They take \SI{66}{\nano\second}, \SI{10}{\nano\second}, and \SI{2.8}{\nano\second}, respectively; thus one cycle is \SI{78.8}{\nano\second}. In contrast, the entropy generation rate of our RHS-TRNG is nearly $24\times$ higher. Although the entropy generation rate can be improved by parallelizing TRNG cells, the use of capacitors as the excitation source of MTJ limits its scalability to  8 bits.

Amirany et al.\cite{amirany2020true} designed a TRNG based on a neuromorphic variation-tolerant spintronic structure. It uses 4 MTJs to control the generation of a random bit and uses XNOR as a post-processing circuit to ensure the stability of the output probability. The proposed neuromorphic spin-based TRNG has typical reset-write-read three phases, where the first two phases take \SI{5}{\nano\second} and \SI{10}{\nano\second}, respectively. So the entropy generation rate of our work is about 4.5x higher, and a longer MTJ  lifetime is obtained due to less writing.

Qu et al. proposed two variation-tolerant TRNG designs, which are referred to as parallel design\cite{qu2017true} and MTJ-pair design\cite{qu2018variation} hereafter. The former utilizes a parallel structure to mitigate the variation effect, while the latter leverages the symmetry of two MTJs to eliminate correlation. We compared these two designs with our proposed RHS-TRNG, and evaluated their abilities to cope with PVT variations under consistent parameters and simulation conditions. For consistency, we configured the parallel design with two MTJs in parallel, adding a set of 16 MTJs in parallel configurations in the evaluation process variation. The voltage and temperature tolerance of the three designs is shown in Fig.~\ref{compare}. It can be observed that the parallel design exhibits insufficient resilience to voltage and temperature variations, as the MTJs configured in parallel are subject to the same variations under both conditions. Fig.~\ref{compare_p} presents a comparison of the three designs under process variations. When only two MTJs are used, the parallel design exhibits a Shannon entropy of only 0.88 and a minimum entropy as low as 0.51 for outputting random sequence under process variations, failing to tolerate process variations. Only by scaling up to 16 MTJs does it achieve a Shannon entropy above 0.99 and a minimum entropy of 0.96, which is slightly lower than our design. The MTJ-pair design can effectively handle all three variations, but its mechanism may lead to errors (such as switching of both MTJs), resulting in lower minimum entropy compared to our design. The generation rates of the two designs are 66.7-177.8 Mb/s and 66.7 Mb/s, respectively, which are much lower than our design (303 Mb/s).

Table.~\ref{tab4} compares RHS-TRNG with five representative MTJ-based TRNGs in the literature at three aspects: throughout, energy consumption, and area, and ranks them in ascending order of throughput rate. As can be seen, RHS-TRNG provides the highest throughout, while having acceptable area and energy consumption.

\begin{table}[t]
	\caption{Comparison of RHS-TRNG and other MTJ-based TRNG designs.}
    \label{tab4}
	\vspace{-20pt}
	\begin{center}
		\resizebox{0.5\textwidth}{!}{
			\bgroup
			\def\arraystretch{1.2}%  1 is the default, change whatever you need
			\begin{tabular}{c|c|c|c}		
          \textbf{ TRNG} & \textbf{Throughput} & 	\textbf{Energy Consump} & \textbf{Area}  \\ & (\SI{}{Mb/s})& (\SI{}{\pico\joule/bit})&(\SI{}{\micro\metre^2})\\ \hline \hline

         \cite{vodenicarevic2017low} & 0.5-100$\times10^{-3}$ & 2-20$\times 10^{-3}$&2\\\hline 
          
         \cite{perach2019asynchronous} & 7.7-15.1  &5.7-13.4  &50.6-200.6  \\ \hline 
         
         \cite{amirany2020true} & 50&1.1  &219 \\\hline

         \cite{qu2017true}   & 66.7-177.8     & 0.6-0.8   &3.8-7.6\\\hline 
         \cite{qu2018variation}   & 66.7     & 0.8   &3.84\\\hline 
               
          This work  &  303 &  2.65-5.3 &  14.5-24.29   \\ \hline   
           
			\end{tabular} 
			\egroup
		}
	\end{center}%
%	\vspace{-10pt}%
\end{table}

\vspace*{-0.8\baselineskip} 
\section{Discussion}
\label{secdis}

In this section, we discuss additional topics of interest for future research work. First, emerging spin devices such as VCMA and SOT also have the potential to serve as TRNG entropy sources while offering  lower power consumption and switching overheads. Nevertheless, ensuring their reliability may require more complex device and circuit designs, which are topics worth investigating.
Second, our design enables the integration of integer and floating-point TRNGs into computing systems. We evaluated the integer TRNG for applications, but the floating-point design compromised precision to some extent. It is worth exploring how to provide high-precision floating-point random numbers for applications that demand such accuracy.
Third, TRNGs can offer potential benefits such as high throughput and statistical quality. We evaluated the effect of high throughput on applications using option pricing as an example. However, integrating TRNGs into systems to improve statistical quality for applications such as cryptography still poses a challenge and will be an interesting research direction for future work.

\vspace*{-0.8\baselineskip} 
\section{Conclusion}
\label{sec7}
In this paper, we have presented  a self-stabilized STT-MTJ-based TRNG: RHS-TRNG. It generates high-quality random bit sequences at the maximum speed of \SI{303}{Mb/s}, which is higher than all prior works to the best of our knowledge. By exploiting cell-level parallelism, higher random bit throughput can be supplied depending on the need of target applications. RHS-TRNG  not only exhibits a strong immunity against PVT variations but only has a longer lifetime, thanks to our circuit design with bidirectional switching currents and dual generator units. We have also integrated RHS-TRNG into a RISC-V processor and demonstrated   that it can significantly accelerate programs that have a strong demand for random numbers, e.g., the Monte Carlo option pricing program. With seamless circuit/system co-design, this work also demonstrates that  Spintronics can be a great driving force to further boost computing system performance in the post-Moore era.

\vfill
\end{document}